\documentclass[lettersize,journal]{IEEEtran}
\usepackage{amsmath,amssymb,amsfonts}
\usepackage{algorithmic}
\usepackage{mathtools}
\usepackage{algorithm}
\usepackage{array}
\usepackage[font=normalsize]{subfig}
\usepackage{textcomp}
\usepackage{stfloats}
\usepackage{url}
\usepackage{verbatim}
\usepackage{siunitx}
\usepackage{graphicx}
\usepackage{cite}
\usepackage{makecell}
\usepackage[switch]{lineno} 
\usepackage{booktabs}

\usepackage[dvipsnames]{xcolor}
\usepackage{placeins}
\usepackage{xifthen}
\usepackage{colortbl}
\usepackage{bbm}

\usepackage{array}
\newcolumntype{H}{>{\setbox0=\hbox\bgroup}c<{\egroup}@{}}
\newcommand{\prob}[1][]{
\ifthenelse{\isempty{#1}}%
      {\ensuremath{P}}%
    {\ensuremath{P\left\(#1\right\)}}%
}
\newcommand{\vect}[1]{\boldsymbol{\mathrm{#1}}}
\newcommand{\mat}[1]{\boldsymbol{\mathrm{#1}}}

\newcommand{\diag}[1]{\mathrm{diag}\left(#1\right)}

\newcommand{\expt}[1]{\mathbb{E}\left(#1\right)}

\DeclareMathOperator*{\argmax}{arg\,max}

\newcommand{\review}[1]{{{#1}}}

\usepackage{pgfplots}
\usepackage{nicefrac}
\usepgfplotslibrary{groupplots,dateplot}

\pgfplotsset{compat=1.18}
\usepgfplotslibrary{fillbetween}
\usepackage{tikz}
\usetikzlibrary{positioning}
\usepackage{hyperref}
\usepackage{acronym}
\usepackage[capitalize]{cleveref}
\usepackage{subfig}
\usetikzlibrary{patterns}
\usepgfplotslibrary{polar}

\usetikzlibrary{shapes}
\usepackage{pgfplotstable}
\usepackage{filecontents}

\usepackage[acronym]{glossaries}
\newacronym[plural=BSs,firstplural=base stations (BSs)]{bs}{BS}{base station}
\newacronym[plural=PAs,firstplural=power amplifiers (PAs)]{pa}{PA}{power amplifier}
\newacronym{sdr}{SDR}{signal-to-distortion ratio}
\newacronym{snr}{SNR}{signal-to-noise ratio}
\newacronym{sndr}{SNDR}{signal-to-noise-and-distortion ratio}
\newacronym{snidr}{SNIDR}{signal-to-noise-and-interference-and-distortion ratio}
\newacronym{los}{LOS}{line-of-sight}
\newacronym{z3ro}{Z3RO}{zero third-order distortion}
\newacronym{mimo}{MIMO}{multiple input multiple output}
\newacronym{mrt}{MRT}{maximum ratio transmission}
\newacronym{dpd}{DPD}{digital pre-distortion}

\newacronym{ula}{ULA}{uniform linear array}

\newacronym{psd}{PSD}{power spectral density}
\newacronym{aclr}{ACLR}{adjacent channel leakage ratio}
\newacronym{cpu}{CPU}{central processing unit}
\newacronym{iid}{i.i.d.}{independently and identically distributed}
\newacronym{ue}{UE}{user equipment}
\newacronym{nlos}{NLoS}{non-line-of-sight}
\newacronym{hpbm}{HPBM}{half power beam width}
\newacronym{rts}{RTS}{ray tracing simulator}
\newacronym{ag}{AG}{array gain}
\newacronym{arp}{ARP}{Antenna Reference Point}
\newacronym{ecdf}{eCDF}{empirical cumulative distribution function}
\newacronym{evm}{EVM}{Error Vector Magnitude}
\newacronym{fft}{FFT}{fast Fourier transform}
\newacronym{ib}{IB}{in-band}
\newacronym{im}{IM}{intermodulation}
\newacronym{mf}{MF}{matched filterifng}
\newacronym{mr}{MR}{maximum ratio}
\newacronym{ofdm}{OFDM}{orthogonal frequency division multiplexing}
\newacronym{oob}{OOB}{out-of-band}
\newacronym{papr}{PAPR}{peak-to-average power ratio}
\newacronym{rx}{RX}{Receiver}
\newacronym{trp}{TRP}{Transmission Reception Point}
\newacronym{tx}{TX}{transmitter}
\newacronym{zf}{ZF}{zero forcing}
\newacronym{dl}{DL}{downlink}
\newacronym{rzf}{RZF}{regularized zero forcing}
\newacronym{slc}{SLC}{spatial leakage suppression}
\newacronym{kpi}{KPI}{key performance indicator}
\newacronym{awgn}{AWGN}{additive white Gaussian noise}
\newacronym{qam}{QAM}{quadrature amplitude modulation}
\newacronym{amam}{AM/AM}{amplitude modulation to amplitude modulation}
\newacronym{ampm}{AM/PM}{amplitude modulation to phase modulation}
\newacronym{zmcscg}{ZMCSCG}{zero mean circularly symmetric complex Gaussian}

\newacronym{gnn}{GNN}{graph neural network}
\newacronym{ccnn}{CCNN}{circular convolutional neural network}
\newacronym{lrelu}{LReLU}{leaky rectified linear unit }
\newacronym{sdg}{SDG}{Sustainable Development Goal}
\newacronym{ibo}{IBO}{input back-off}
\newacronym{cdf}{CDF}{cumulative distribution function}
\newacronym{nn}{NN}{neural network}
\newacronym{mlp}{MLP}{multilayer perceptron}
\newacronym{mmimo}{mMIMO}{massive MIMO}
\newacronym{dab}{DAB}{distortion-aware beamforming}
\newacronym{flops}{FLOPs}{floating point operations}
\newacronym{mmwave}{mmWave}{millimeter-wave}
\newacronym{gops}{GFLOPs/s}{giga floating point operations per second}
\newacronym{sp}{SP}{single precision}
\newacronym{dsp}{DSP}{digital signal processing}
\newacronym{dac}{DAC}{digital-to-analog converter}
\newacronym{aqnm}{AQNM}{additive quantization noise model}
\newacronym{sinqdr}{SINQDR}{signal-to-interference-noise-and-quantization-distortion ratio}
\newacronym{nmsqe}{NMSQE}{normalized mean squared quantization error}
\newacronym{mmsqe}{MMSQE}{minimum mean squared quantization error}
\newacronym{agc}{AGC}{automatic gain control}
\newacronym{rf}{RF}{radio frequency}
\newacronym{sfdr}{SFDR}{spurious free dynamic range}
\newacronym{adc}{ADC}{analog-to-digital converter}
\newacronym{mmse}{MMSE}{minimum mean squared error}
\newacronym{cnn}{CNN}{convolutional neural network}
\newacronym{mse}{MSE}{mean squared error}
\newacronym{nmse}{NMSE}{normalized mean squared error}
\newacronym{nco}{NCO}{neural combinatorial optimization}
\newacronym{rfdac}{RF-DAC}{radio frequency digital-to-analog converter}

\glsdisablehyper

\usepackage{amsthm}
\newtheoremstyle{mystyle}
  {}
  {}
  {\itshape}
  {}
  {\bfseries}
  {.}
  { }
  {}
\theoremstyle{mystyle}

\begin{document}

\title{Learning to Quantize and Precode in Massive MIMO Systems for Energy Reduction: a Graph Neural Network Approach
}

\author{Thomas Feys, Liesbet Van der Perre,
        François Rottenberg
\thanks{The authors are with ESAT-DRAMCO, Campus Ghent, KU Leuven, Belgium (email: \href{mailto:thomas.feys@kuleuven.be}{thomas.feys@kuleuven.be}).}
\thanks{We thank NVIDIA for providing the GPU that accelerated our simulations.} \thanks{Partially funded by 6GTandem part of SNS JU under the European Union’s Horizon Europe research and innovation program, Grant 101096302.} \thanks{This work is Co-funded by the European Union under Grant Agreement 101191936. Views and opinions expressed are however those of the author(s) only and do not
necessarily reflect those of all 6GTandem/SUSTAIN-6G consortium parties nor those of the
European Union or the SNS-JU (granting authority). Neither the European Union nor
the granting authority can be held responsible for them. }}



\maketitle

\begin{abstract}

Massive MIMO systems are moving toward increased numbers of radio frequency chains, higher carrier frequencies and larger bandwidths. As such, \glspl{dac} are becoming a bottleneck in terms of hardware complexity and power consumption. In this work, non-linear precoding for coarsely quantized downlink massive MIMO is studied. Given the NP-hard nature of this problem, a \gls{gnn} is proposed that directly outputs the precoded quantized vector based on the channel matrix and the intended transmit symbols. The model is trained in a self-supervised manner, by directly maximizing the achievable rate. To overcome the non-differentiability of the objective function, introduced due to the non-differentiable \gls{dac} functions, a straight-through Gumbel-softmax estimation of the gradient is proposed. The proposed method achieves a significant increase in achievable sum rate under coarse quantization. For instance, in the single-user case, the proposed method can achieve the same sum rate as \gls{mrt} by using one-bit \glspl{dac} as compared to 3 bits for \gls{mrt}. \review{This reduces the \gls{dac}'s power consumption by a factor 4-7 and 3 for baseband and RF \glspl{dac} respectively}. This, however, comes at the cost of increased digital signal processing power consumption. \review{When accounting for this, the reduction in overall power consumption holds for a system bandwidth up to \SI{3.5}{\mega \hertz} for baseband \glspl{dac}, while the RF \glspl{dac} can maintain a power reduction of 2.9 for higher bandwidths}.  Notably, indirect effects, which further reduce the power consumption, such as a reduced fronthaul consumption and reductions in other components, are not considered in this analysis. 

\end{abstract}


\section{Introduction}
\label{sec:intro}
\subsection{Problem Statement}
Massive \gls{mimo} is a key enabler for 5G and is expected to be further developed for the next generation~\cite{björnson2019massivemimoreality}. Massive \gls{mimo} leverages many antennas at the \gls{bs} to spatially focus the signal to the user locations. This increases the achievable data rates and the spectral efficiency~\cite{mimo_for_next_gen}. While the capacity is increased, this does incur a significant hardware cost. In a fully digital massive \gls{mimo} system, each radio frequency chain is typically equipped with its own, \gls{adc}, \gls{dac}, \gls{pa} and antenna, among others. 

\review{The use of many radio frequency chains increases both the hardware complexity and the power consumption. To combat this, in this work we focus on reducing the bit-width of \glspl{dac}. This is motivated by the current move towards higher carrier frequencies and higher bandwidths~\cite{thz}, which further increases the complexity and the power consumption of the \glspl{dac}. The reason for this increased power consumption is threefold. First, the power consumption of typical current steering \glspl{dac} scales exponentially with the number of bits and linearly with the sampling frequency~\cite{pwr_dac}. This raises concerns given the ever-increasing bandwidth of wireless systems. Second, \glspl{rfdac} are becoming increasingly popular in wireless systems due to their ability to directly synthesize the \gls{rf} signal at the desired carrier frequency, which reduces analog complexity~\cite{rfsoc_book}. This is possible by relying on a high oversampling factor to place the signal at the desired carrier frequency, bypassing traditional analog upconversion. However, due to the increased oversampling factor, coupled with the current move towards higher carrier frequencies and higher bandwidths, the power consumption of these \glspl{rfdac} is significantly higher as compared to traditional \glspl{dac}. To offset this increased power consumption, we investigate how the bit-width can be reduced, providing reduced power consumption in the \glspl{dac}, while maintaining system performance. Finally, the spurious free dynamic range, that is the frequency range over which the \gls{dac} operates free from unwanted spurious signals, is limited by the resolution of the \gls{dac}~\cite{sfdr_dac}. Hence, high resolution \glspl{dac} become problematic at higher carrier frequencies and bandwidths, as additional nonlinearities are introduced. Next to this, decreasing the bit-width can have significant indirect benefits. For instance, the power consumption of the digital fronthaul, which connects the baseband unit with the antenna array, can be significantly reduced when fewer bits are considered. Alternatively, if the bit-width reduction can be propagated into the baseband processing this would have a big impact on its power consumption.}

 Consequently, low-resolution \glspl{dac} are vital to overcome these hardware limitations and reduce the power consumed in the many \glspl{dac}. However, by using lower resolution \glspl{dac} a significant quantization error is introduced, which limits the \gls{sdr} and the achievable data rates. To overcome this, we present an approach to limit this quantization distortion by leveraging the degrees of freedom available from the many antennas at the \gls{bs}. This is done by proposing a \gls{nn}-based non-linear precoding scheme that mitigates this distortion. By doing so, the \glspl{dac} can utilize fewer bits, while a high achievable data rate can be maintained in the network.

\subsection{State-of-the-Art Quantization in Massive MIMO}
Quantization in Massive \gls{mimo} systems has been widely studied. A rich body of literature focuses on the effects of \glspl{adc} at the receiver~\cite{adc_rate, adc_rate2, mixed_adc_rate, adc_rician, adc_processing}. 
In this work, we focus on the effects of \gls{dac} distortion. Two trends have emerged when dealing with \gls{dac} distortion in \gls{mimo} systems. On the one hand, there is a focus on 1-bit systems, largely driven by their analytical tractability and simplified hardware design. On the other hand, several works tackle the more general few-bit case. However, due to the intractability, simplifying assumptions have been made which are only valid when many users or many bits are considered. As shown in this work, distortion due to the \glspl{dac} is most limiting when few users are present. This is because distortion becomes more spatially spread out as the number of users rises, similar to the case of non-linear \gls{pa} distortion~\cite{distortion_beamformed2}. Hence, there is a need for a solution to the quantization-aware precoding problem for the case of few users and few bits. 

\subsubsection{1-bit DACs}
In~\cite{1bit_zf}, a 1-bit quantized \gls{zf} precoder is studied. It is shown that when the number of transmit antennas goes to infinity, the received signals are scaled versions of the desired symbols. Next to this, in~\cite{mmse_1bit} a two-stage digital-analog precoder is proposed that takes into account one-bit \glspl{dac} at the transmitter and one-bit \gls{adc} at the receiver. Additionally, in~\cite{jacobsson_quantized_2017, jacobsson_1bit_nonlin} several one-bit non-linear precoders are proposed i) based on a semidefinite relaxation leading to a convex problem, ii) based on a squared $l_{\infty}$-norm relaxation which also leads to a convex problem and iii) based on sphere decoding rather than solving the non-linear quantized precoding problem, which essentially is an NP-hard closest-vector problem, through exhaustive search.

\subsubsection{Few-bit DACs}
In~\cite{transmit_processing_dacs}, \gls{mmse} linear processing is proposed where 4-6 bit \glspl{dac} are deployed at the \gls{bs}. However, this work relies on the high-resolution assumption to specify the quantization error, i.e., when the number of bits $b$ is large, the quantization error of an optimal non-uniform quantizer approximately scales with $2^{-2b}$. The work in~\cite{EE_Optimization} optimizes the energy efficiency while taking into account the power consumption and distortion due to low-resolution \glspl{adc} and \glspl{dac}. However, the distortion covariance matrix is assumed to be diagonal (i.e., uncorrelated distortion). This assumption holds when many users with an \gls{iid} Rayleigh fading channel are considered~\cite{hardware_correlation}. Hence, this assumption does not hold in the most critical case for quantization distortion, i.e., the few user-case. Similarly, in~\cite{EE_mmwave} the quantized single-user \gls{mimo} case is studied. The energy efficiency of digital and hybrid precoders is evaluated. However, this study also relies on the assumption of uncorrelated distortion, which only holds when many users are present.

In summary, there is a need for transmit processing that addresses the case of few-bit \glspl{dac} when few users are present. In this study, we overcome this gap by proposing neural network-based non-linear precoders that are trained for both the one-bit and few-bit \gls{dac} cases, without relying on assumptions that only hold for high-resolution \glspl{dac} or a high number of users.

\subsection{State-of-the-Art Neural Network-based Precoding}
\glspl{nn} have become an interesting approach for learning to precode. Some works focus on learning \gls{nn}-based precoders as a low-complexity alternative to high-complexity conventional precoding algorithms~\cite{fast_nn_beamforming, model_driven_beamforming, feys_unfolding}. Furthermore, \gls{nn}-based precoding has been used to solve precoding problems that do not have a simple solution and where conventional algorithms show limited performance~\cite{model_driven_beamforming, gnn_feys, icc_ccnn}. In these works a wide range of \gls{nn} architectures is used ranging from \glspl{mlp}, \glspl{cnn}, transformers and \glspl{gnn}. Each of these architectures has a certain structure that limits the types of functions these \glspl{nn} can learn. In general, the aim of selecting a certain architecture is to reduce the hypothesis space of the \gls{nn}, that is the space of all possible functions the \gls{nn} can learn. Reducing the hypothesis space typically corresponds to a \gls{nn} that has fewer trainable parameters. Consequently, less training data is needed and training becomes easier. However, the cost of reducing the hypothesis space is a less expressive model, which can cause the \gls{nn} to not be able to learn the desired function. However, if one can reduce the hypothesis space in an informed manner such that the desired function is still covered by this reduced hypothesis space, the reduced expressivity of the model does not affect the learning performance~\cite{inductivebias, gnn_feys}. Note that this is closely linked to the bias-variance trade-off~\cite{dl_goodfellow}. A model with a large hypothesis space will have a low bias but high variance, while reducing the hypothesis space leads to lower variance, but might increase the bias. The goal is to find a hypothesis space with an optimal trade-off such that both the bias and variance are minimized. 
Recent works have shown that, for linear precoding, \glspl{gnn} are especially well-suited~\cite{understanding_gnn_precoding, cnn_gnn, gnn_feys}. This is due to their permutation equivariance properties. Simply put, if the order of the rows/columns in the input matrix to the \gls{nn} is permuted, the order of the rows/columns in the output matrix should be permuted accordingly. Concerning the linear precoding problem, if the rows/columns of the channel matrix (input) are permuted, the rows/columns of the precoding matrix (output) should be permuted accordingly. This is an inherent property of linear precoders, which is respected by using \glspl{gnn}. Hence, by using \glspl{gnn} the hypothesis space is reduced to only include permutation equivariant functions, thereby reducing the variance while keeping the bias low.

Concerning quantization-aware precoding, several works have introduced the use of \glspl{nn} for the 1-bit case. In~\cite{1bit_nn_unfolding}, the authors propose to unfold the biconvex 1-bit precoding algorithm from~\cite{jacobsson_vlsi}. 
Next to this, in~\cite{deep_learninng_1bit} non-linear precoding for the 1-bit case is learned in a supervised manner. 

More broadly, the quantized precoding problem can be seen as a combinatorial problem, as the goal is to select the best output level at each \gls{dac} from a finite set. Our work is inspired by a sub-field of deep learning, known as \gls{nco}~\cite{bello2017neuralcombinatorialoptimizationreinforcement, garmendia2022neuralcombinatorialoptimizationnew, ML_CO_bengio}. \Gls{nco} aims at training \glspl{nn} that can solve combinatorial problems. Given the NP-hard nature of many combinatorial problems, it is often hard to obtain optimal solutions that can be used as labels. Hence, supervised learning for combinatorial optimization is often costly or impossible. Consequently, \gls{nco} is typically done in combination with reinforcement learning, where the proposed solution by the \gls{nn} is evaluated. This evaluation is then used as a reward signal to update the weights of the \gls{nn}~\cite{bello2017neuralcombinatorialoptimizationreinforcement}. More recently, one-shot neural combinatorial solvers have been introduced that do not rely on reinforcement learning~\cite{one_shot_comb_solvers}. These methods directly predict the decision variables for the problem. The decision variables are then evaluated using a differentiable objective function. This differentiable objective function is then used as a self-supervised loss function, to update the weights of the \gls{nn} through stochastic gradient descent~\cite{one_shot_comb_solvers}. The benefit of this methodology is \textit{i)} that no labels are required,  \textit{ii)} unsupervised learning typically generalizes better as compared to the mimic learning that happens in the supervised setting, which can be attributed to the unsupervised loss function capturing more underlying physical insights of the problem, and  \textit{iii)} it is empirically more efficient than reinforcement learning~\cite{one_shot_comb_solvers}. The main drawback of this method is that the objective function should be differentiable, which is not always the case. In this work, we follow the one-shot neural combinatorial solver methodology. Given the non-differentiable nature of our objective function (caused by the non-differentiable \gls{dac} functions), we utilize a Gumbel-softmax-based~\cite{gumbel_review} relaxation which renders the objective function differentiable in the backward pass but still discrete in the forward pass.





\subsection{Contributions}

In this work, a novel \gls{gnn}-based non-linear precoding method is proposed that achieves high data rates under coarse quantization in a downlink massive MIMO system. \Cref{sec:problem_forulation} outlines the system model. In~\cref{sec:rad_plots_lin_quant} the critical case in terms of distortion is identified as the few users, few bits case. Hence, the proposed precoder is developed considering few users and few bits, i.e., it does not rely on assumptions of uncorrelated distortion that are only valid when many users are present or many bits are considered, as is often done in the current literature. In~\cref{sec:nn_for_prec}, based on the permutation equivariance properties of the non-linear precoding problem a \gls{gnn} is proposed that captures these properties. Additionally, an unsupervised training procedure is outlined to train this \gls{gnn}. This is especially challenging given the non-differentiable nature of the objective function, caused by the non-differentiable \gls{dac} functions. To overcome this we propose a Gumbel-softmax-based relaxation of the objective function. In~\cref{sec:complexity} the complexity of the proposed \gls{gnn} is computed. \Cref{sec:results} provides extensive simulation results to validate the performance of the proposed \gls{gnn}. It is shown that the proposed precoder can achieve a significant increase in terms of achievable sum rate. Radiation patterns illustrate that this gain is achieved by transmitting the distortion in non-user directions. In \cref{sec:power_cons} the reduction in power consumption of the \glspl{dac} is computed while taking into account the additional processing power consumption of the \gls{gnn}. When solely considering the \glspl{dac} consumption, it is shown that the proposed method reduces the power consumption by a factor of 4-7. However, when taking the \gls{gnn} processing power consumption into account the power reduction is only achieved for bandwidths up to \SI{3.5}{\mega\hertz}, without taking into account indirect effects such as reduced fronthaul consumption. Finally,~\cref{sec:conclusion} concludes the paper.

\subsection*{Notations}
 Vectors and matrices are denoted by bold lowercase and bold uppercase letters respectively. A vector function is denoted by a bold letter while a scalar function is denoted by a non-bold letter. Superscripts $(\cdot)^*$, $(\cdot)^{\intercal}$ and $(\cdot)^{H}$ stand for the conjugate, transpose and Hermitian transpose operators respectively. The expectation is $\mathbb{E}(.)$. The $M\times M$ identity matrix is given by $\mat{I}_M$. The main diagonal of a square matrix $\mat{A}$ is given by $\diag{\mat{A}}$. The trace of a matrix is given by $\mathrm{Tr}\left(\cdot\right)$. The element at location $(i,j)$ in matrix $\mat{A}$ is indicated as $[\mat{A}]_{i,j}$. Similarly, element $i$ of vector $\vect{a}$ is denoted as $[\vect{a}]_i$ or in scalar form as $a_i$. $\mathcal{G} = (\mathcal{V}, \mathcal{E})$ denotes a graph where $\mathcal{V}$ is the set of nodes and $\mathcal{E}$ the set of edges. The edge going from node $a \in \mathcal{V}$ to node $b \in \mathcal{V}$ is denoted as $(a, b) \in \mathcal{E}$. The neighborhood of a node $a$ is denoted as $\mathcal{N}(a) = \{b \in \mathcal{V}: (a, b) \in \mathcal{E}\}$. The indicator function denotes one-hot encoded vectors where $\mathbbm{1}_i^{N}$ is a length $N$ vector containing zeros except for a one at index $i$.

\section{System Model}

\begin{figure*}
\centering
\begin{minipage}[b]{.48\textwidth}
    \includegraphics[width=1\linewidth]{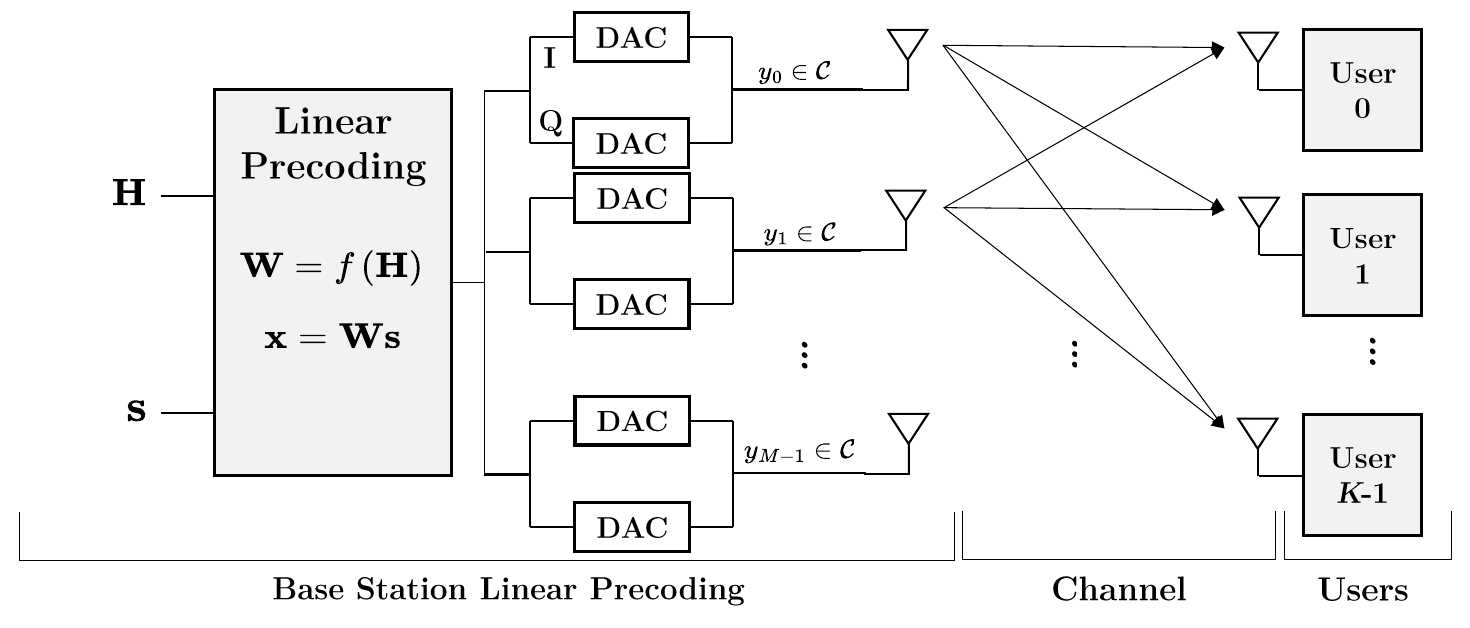}
\caption{Simplified system overview of linear quantized precoding.}\label{fig:overview_lin}
\end{minipage}\qquad
\begin{minipage}[b]{.48\textwidth}
    \includegraphics[width=0.9\linewidth]{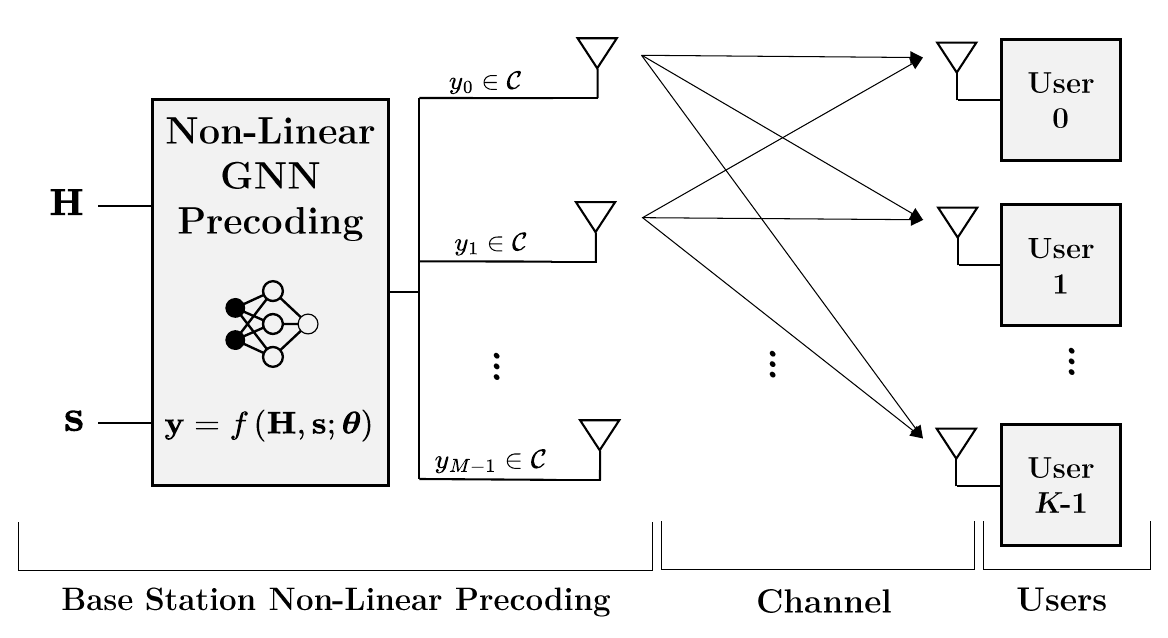}
\caption{Simplified system overview of non-linear quantized precoding.}\label{fig:overview_nonlin}
\end{minipage}
\end{figure*}


\label{sec:problem_forulation}
\subsection{Linear Quantized Precoding} 
\label{sec:lin_quant_prec}

Consider a \gls{bs} with $M$ transmit antennas and $K$ single antenna users, operating in the downlink. When considering digital full-precision linear precoding, the precoded vector is obtained as $\vect{x} = \mat{W} \vect{s}$, where $\vect{x} \in \mathbb{C}^{M}$ are the precoded symbols, $\mat{W} \in \mathbb{C}^{M \times K}$ is the precoding matrix and $\vect{s} \in \mathbb{C}^{K}$ is the symbol vector, where the symbols between different users are uncorrelated and zero mean. When a single user is present we consider \gls{mrt} precoding, with precoding matrix $\mat{W}^{\mathrm{MRT}} = \alpha_{\mathrm{norm}} \mat{H}^*$. When multiple users are present \gls{zf} precoding is considered as $\mat{W}^{\mathrm{ZF}} = \alpha_{\mathrm{norm}} \mat{H}^* (\mat{H}^{\intercal} \mat{H}^*)^{-1}$.
Both for \gls{mrt} and \gls{zf} precoding the power normalization constant is defined as $\alpha_{\mathrm{norm}} = \sqrt{P_T/\mathrm{Tr}(\mat{W} \mat{W}^H)}$ so that the total transmit power is normalized to $P_T$. After precoding, the digital baseband signal $\vect{x}$ is quantized. This is done for the real and imaginary dimensions per antenna
\begin{equation}
\begin{aligned}
	\vect{y} = \boldsymbol{\mathcal{Q}}(\vect{x}) 
   &= [\mathcal{Q}(\Re\{x_0\}) + \jmath \mathcal{Q}(\Im\{x_0\}),  \cdots,  \\
    & \mathcal{Q}(\Re\{x_{M-1}\}) + \jmath \mathcal{Q}(\Im\{x_{M-1}\}) ]^{\intercal}
\end{aligned}
\end{equation}
where $\boldsymbol{\mathcal{Q}}(\cdot): \mathbb{C}^{M} \mapsto \mathcal{C}^M$ denotes the quantization function which characterizes the combined operation of the $2M$ \glspl{dac} at the \gls{bs} and $\mathcal{Q}(\cdot): \mathbb{R} \mapsto \mathcal{L}$ denotes the real-valued quantizer. These quantizers are further detailed in~\cref{sec:quant}. 

\review{
This quantization can be characterized by the \gls{aqnm}, originally proposed by~\cite{og_aqnm}, which is a special case of the Bussgang decomposition with the nonlinear function being a quantizer that represents each quantization interval by its mean value~\cite{demir2020bussgang}. In this work, we follow the convention of the Bussgang decomposition
\begin{align}
            \vect{y} = \boldsymbol{\mathcal{Q}}(\vect{x}) = \vect{\Phi}_{\alpha}  \vect{x} + \vect{q} \label{eq:aqnm}
\end{align}
where the distortion $\vect{q}$ is uncorrelated with $\vect{x}$, i.e., $\expt{\vect{xq}^H} = \mat{0}$.
The Bussgang matrix is given by $\mat{\Phi}_{\alpha} = \diag{\alpha_0, \cdots, \alpha_{M-1}}  \in \mathbb{R}^{M\times M}$, where $\alpha_m = \frac{\expt{y_m x_m^*}}{\expt{|x_m|^2}}$. The link with the \gls{aqnm} can be made by defining the \gls{nmsqe} after the $m^{\mathrm{th}}$ \gls{dac} as
\begin{align}
    \beta_m = \frac{\expt{(x_m- y_m)(x_m-y_m)^*}}{\expt{|x_m|^2}}
\end{align}
and relating the Bussgang gain to the \gls{nmsqe} after each \gls{dac} as $\mat{\Phi}_{\alpha} = \mat{I}_M - \mat{\Phi}_{\beta}$, where $\mat{\Phi}_{\beta} = \diag{\beta_0, \cdots, \beta_{M-1}}  \in \mathbb{R}^{M\times M}$.
}


\subsection{Quantization}\label{sec:quant}
\subsubsection{Quantization of a Complex-Valued Vector}
The precoded (unquantized) vector is given by $\vect{x} \in \mathbb{C}^{M}$. After quantization, the precoded vector is given by $\vect{y} \in \mathcal{C}^{M}$, where the set $\mathcal{C}$ is the set of complex output levels/codebook. When considering infinite precision the set $\mathcal{C}$ coincides with $\mathbb{C}$. However, due to the use of \glspl{dac}, the set $\mathcal{C}$ has a finite cardinality. The set of real-valued \gls{dac} outputs is given by $\mathcal{L} = \{l_0, \cdots l_{L-1}\}\subset \mathbb{R}$, where $l_i$ are the output values/levels and $L=|\mathcal{L}| = 2^b$ is the number of quantization levels with $b$ the number of considered bits per real dimension. For each \gls{bs} antenna the same output set is considered for the real and imaginary parts. The set of complex-valued \gls{dac} outputs is then given by $\mathcal{C} = \mathcal{L} \times \mathcal{L}$. This can be seen as applying one real-valued \gls{dac} to the real part of the precoded symbol at each antenna and one real-valued \gls{dac} to the imaginary part. The real-valued quantizer partitions the real line $\mathbb{R}$ into $L$ cells $\mathcal{R}_i$ for $i=0, 1, \cdots, L-1$. The $i$th cell is defined as $\mathcal{R}_i = \{x \in \mathbb{R}: \mathcal{Q}(x) = l_i\}$~\cite{vector_quantization}. From this definition, it follows that $\bigcup_i \mathcal{R}_i = \mathbb{R}$ and $\mathcal{R}_i \cap \mathcal{R}_j = \emptyset$. Additionally, each cell is fully described by its boundary points/thresholds $\mathcal{T} = {\tau_0, \cdots, \tau_L}$,  $\mathcal{R}_i = (\tau_{i}, \tau_{i+1}]$ for $i=0, \cdots, L-1$~\cite{vector_quantization}. For quantizers operating on unbounded inputs typically $\tau_0 = -\infty$ and $\tau_{L}=+\infty$. The optimal values of the output labels and thresholds depend on the type of quantization used and the input distribution, as described in the next section.


\subsubsection{Non-Uniform Quantization} \label{sec:non-uniform-quant}
The input distribution in this work is Gaussian, which is a valid assumption for \gls{ofdm} systems~\cite{ofdm}. For a real-valued Gaussian input, uniform quantization is not optimal in the \gls{mmsqe} sense~\cite{vector_quantization}. A \gls{mmsqe} non-uniform quantizer can be found by solving $\min_{\mathcal{L}, \mathcal{T} } \  \expt{(x-\mathcal{Q}(x))^2} = \sum_{i=1}^N \int_{\mathcal{R}_i} (x-l_i)^2 f_{X}(x) dx$. The Max-Lloyd algorithm as described in~\cite{vector_quantization, max_min_dist} can be used to find the optimal output levels and thresholds. This is a hill-climbing algorithm, which can lead to local optima in the solution, hence it is advised to run the algorithm for several initialization points. This is done to obtain the output levels and thresholds used in this work. 

\subsubsection{Input Normalization}
To achieve optimal quantization the input distribution to the \gls{dac} should be $\mathcal{CN}(0, 1)$. The transmit symbols are $s_k \sim \mathcal{CN}(0,1)$. However, the input to the \gls{dac} is not the transmit symbol but the precoded symbol $x_m = \sum_k w_{m, k} s_k = \vect{w}_m^{\intercal} \vect{s}$. As such, the precoded symbol is a scaled random variable that is complex normal distributed. Its power (variance) can be computed as $\expt{|x_m|^2} = \|\vect{w}_m\|_2^2$.
To adjust the power of the input signal (per antenna chain) to the dynamic range of the DAC, an input normalization\footnote{This is similar to the automatic gain control typically applied at the receiver before the \gls{adc}.} scales the input as $\tilde{x}_m = \frac{1}{\sqrt{\rho_m}} x_m$, where $\rho_m$ is a scalar normalization factor. 
If $\rho_m =   \|\vect{w}_m\|_2^2$ the signal power is rescaled to one.
This normalized signal is sent to the DAC to be quantized as $\tilde{y}_m = \mathcal{Q}(\tilde{x}_m)$.
After quantizing this normalized signal, the output of the DAC should be denormalized as $ y_m = \sqrt{\rho_m} \tilde{y}_m$. This denormalization can be taken into account by the amplification stage, while the normalization can be performed in the digital signal processing.


\subsection{Non-Linear Neural Network-based Quantized Precoding}
\Gls{nn}-based non-linear precoding is considered. This entails that a \gls{nn} is used to learn a mapping from the channel matrix and symbol vector to the precoded quantized symbol vector. The function we want to learn can be represented as
\begin{align}
	\vect{y}_{\mathrm{NL}} = f\left(\mat{H}, \vect{s}; \vect{\theta}\right) \label{eq:nonlin_prec}
\end{align}
where $f(\cdot, \cdot): \mathbb{C}^{M\times K} \times \mathbb{C}^{K} \mapsto \mathcal{C}^{M}$ represents the \gls{nn}, i.e., a learned non-linear function mapping parametrized by its trainable weights $\vect{\theta}$. Given that there is no explicit quantization step for non-linear precoding, the Busgang decomposition with respect to $\vect{x}$ can no longer be computed. More generally, we define the Bussgang decomposition between the precoded quantized signal $\vect{y}$ and the intended transmit symbols $\vect{s}$ as 
\begin{align}
    \vect{y} = \mat{G} \vect{s} + \vect{q} \label{eq:bus_s}
\end{align}
where the Bussgang gain matrix is no longer a square diagonal matrix, it is now defined as $\mat{G} = \expt{\vect{y} \vect{s}^{H}} (\expt{\vect{s} \vect{s}^H})^{-1} = \expt{\vect{y} \vect{s}^{H}}$. The link with the Bussgang decomposition under linear precoding in \eqref{eq:aqnm} is easily made using $\vect{x}=\mat{W}\vect{s}$, as $\mat{G} =  \mat{\Phi}_{\alpha} \mat{W}$.

\subsection{Achievable Sum Rate}\label{sec:bussgang_rx}
Regardless of which precoding technique is considered, the received signal can be written as
\begin{align}
	\vect{r} = \mat{H}^{\intercal} \vect{y} + \vect{v}
\end{align}
where $\vect{r} \in \mathbb{C}^{K}$ is the received signal vector, $\mat{H}\in\mathbb{C}^{M\times K}$ the channel matrix, $\vect{y}\in\mathcal{C}^{M}$ the precoded and quantized transmit symbols and $\vect{v}\in\mathbb{C}^{K}$ additive white Gaussian noise with covariance matrix $\sigma_v^2 \mat{I}_K$.
Using~\eqref{eq:bus_s}, the received signal for user $k$ can be written as
\begin{align}
    r_k = \underbrace{\vect{h}_k^{\intercal} \vect{g}_k s_k}_{\text{intended signal}} + \underbrace{\sum_{k^{\prime} \neq k}  \vect{h}_k^{\intercal} \vect{g}_{k^{\prime}} s_{k^{\prime}}}_{\text{user interference}} + \underbrace{\vect{h}_k^{\intercal} \vect{q}}_{\text{distortion}} + \underbrace{v_k}_{\text{noise}}.
\end{align}
Hence the \gls{snidr} for user $k$ can be written as
\begin{align} \label{eq:snr_new}
    \mathrm{SNIDR}_k = \frac{ |\vect{h}_k^{\intercal} \vect{g}_k|^2 }{\sum_{k^{\prime} \neq k}  |\vect{h}_k^{\intercal} \vect{g}_{k^{\prime}} |^2  +  \vect{h}_k^{\intercal} \expt{\vect{q} \vect{q}^{H}}  \vect{h}_k^{*} + \sigma_v^2}. 
\end{align}


An achievable sum rate $R_{\mathrm{sum}}$, i.e., a lower bound on the capacity, can be obtained by considering a worst-case scenario, where the noise and quantization distortion are jointly Gaussian distributed and independent from the data symbols
\begin{align}
    R_{\mathrm{sum}} = \sum_{k=0}^{K-1} \log_2\left(1 + \mathrm{SNIDR}_k\right). \label{eq:sumrate}
\end{align}


\subsection{Radiation Patterns}
 In Sections~\ref{sec:rad_plots_lin_quant} and \ref{sec:results} a number of radiation patterns are evaluated. For this, a pure \gls{los} channel $\vect{h}_{\mathrm{los}}(\phi_k)$ is considered, the channel coefficient from antenna $m$ to user $k$ is 
\begin{align}
    h_{m,k} = e^{-\jmath m \frac{2\pi}{\lambda_c} d \cos( \phi_k)}\label{eq:lineofsight}.
\end{align}
Here $ m \frac{2\pi}{\lambda_c} d \cos( \phi_k)$ is the antenna-dependent phase shift when considering a \gls{ula} and a narrowband system, $\lambda_c$ is the carrier frequency, $d=\lambda_c/2$ the antenna spacing and $\phi_k$ the user angle. 
The radiation pattern of the intended signal in an arbitrary direction $\tilde{\phi}$ can be written as
\begin{align}
    P_{\mathrm{lin}}(\tilde{\phi}) = \expt{| \vect{h}^{\intercal}_{\mathrm{los}}(\tilde{\phi}) \mat{G} \vect{s}|^2} = \vect{h}^{\intercal}_{\mathrm{los}}(\tilde{\phi})  \mat{G}  \mat{G}^H   \vect{h}^{*}_{\mathrm{los}}(\tilde{\phi}).\label{eq:rad_lin_zf}
\end{align}
The radiation pattern of the quantization distortion is given by
\begin{align}
    P_{\mathrm{dist}}(\tilde{\phi}) = \expt{| \vect{h}^{\intercal}_{\mathrm{los}}(\tilde{\phi}) \vect{q}|^2} = \vect{h}^{\intercal}_{\mathrm{los}}(\tilde{\phi})  \expt{\vect{q} \vect{q}^H}  \vect{h}^{*}_{\mathrm{los}}(\tilde{\phi}). \label{eq:rad_dist_zf}
\end{align}
The \gls{sdr} radiation pattern is defined as $P_{\mathrm{SDR}}(\tilde{\phi}) = \frac{P_{\mathrm{lin}}(\tilde{\phi})}{P_{\mathrm{dist}}(\tilde{\phi})}$.

\section{Spatial Characteristics of Quantization-Induced Distortion}\label{sec:rad_plots_lin_quant}
When considering linear precoding techniques such as \gls{mrt} and \gls{zf}, followed by a quantization step, the radiation patterns of both the linear and distortion signals can be characterized according to (\ref{eq:rad_lin_zf}) and (\ref{eq:rad_dist_zf}). In this section, these numerically obtained radiation patterns are analyzed in order to identify scenarios where the quantization distortion is most critical. 

\begin{figure*}[ht!]
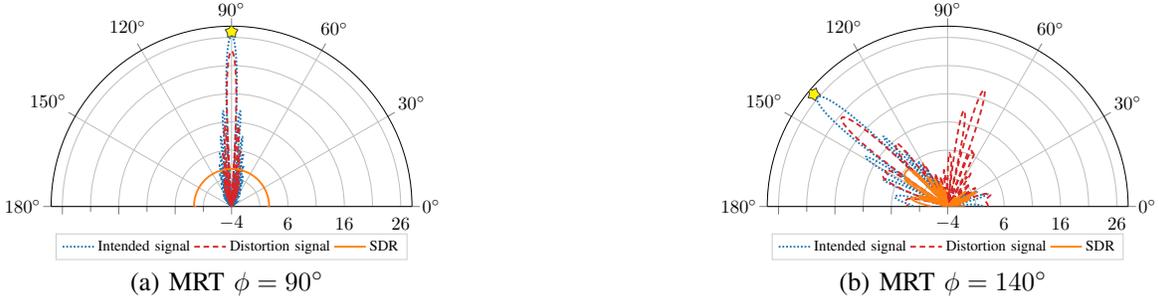

    \centering
    \subfloat[MRT $\phi=90^{\circ}$]{\label{fig:rad_mrt_90}\input{Figs/rad_mrt/rad_mrt_90}}
    \hspace{0.19\linewidth}%
    \subfloat[MRT $\phi=140^{\circ}$]{\label{fig:rad_mrt_140}\input{Figs/rad_mrt/rad_mrt_140}}
	\caption{Radiation pattern of the intended signal $P_{\mathrm{lin}}(\tilde{\phi})$, distortion signal $
P_{\mathrm{dist}}(\tilde{\phi})$ and \gls{sdr} $P_{\mathrm{SDR}}(\tilde{\phi})$ in [dB] for a 1-bit DAC $b=1$,  $M=32$, $K=1$ for MRT precoding for different user angles. Users are indicated by a star. }
	\label{fig:rad_mrt} 
\end{figure*}

 In~\cref{fig:rad_mrt} the radiation pattern of the distortion, the intended signal and \gls{sdr} are depicted for $K=1$, $b=1$ and $M=32$. For some pathological cases (for instance $\phi \in \{0^{\circ}, 60^{\circ}, 90^{\circ}, 120^{\circ}, 180^{\circ}\}$), the beamforming gain assigned to the intended and distortion signal are identical, which leads to an \gls{sdr} that is uniformly distributed in space, as seen in~\cref{fig:rad_mrt_90}. In most other cases, the distortion receives a beamforming gain in multiple directions, with the user direction being the dominant one, as seen in~\cref{fig:rad_mrt_140}. This leads to a small \gls{sdr} in the user direction. Increasing the number of bits generally leads to identical but scaled radiation patterns, where the intended signal increases, the distortion decreases and the \gls{sdr} increases. 

\begin{figure*}[ht!]
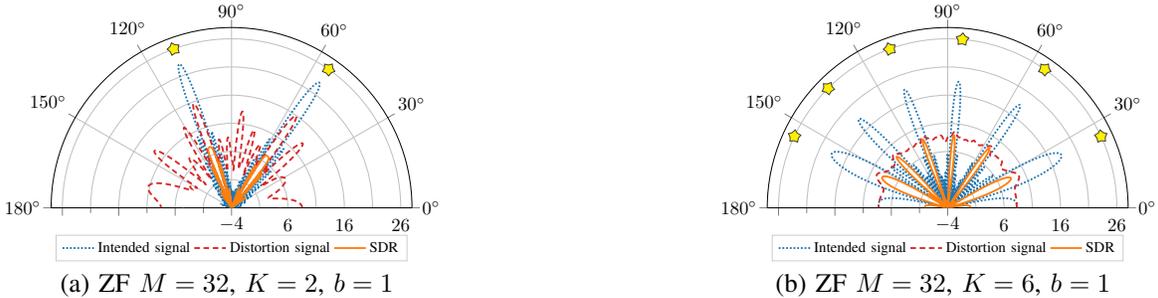

    \centering
    \subfloat[ZF $M=32$, $K=2$, $b=1$]{\label{fig:rad_zf_k2}\input{Figs/rad_zf/rad_zf_k2}}
        \hspace{0.19\linewidth}%
    \subfloat[ZF $M=32$, $K=6$, $b=1$]{\label{fig:rad_zf_k6}\input{Figs/rad_zf/rad_zf_k6}}
	\caption{Radiation pattern for \gls{zf} of the intended signal $P_{\mathrm{lin}}(\Tilde{\phi})$, distortion $P_{\mathrm{dist}}(\Tilde{\phi})$ and \gls{sdr} $P_{\mathrm{SDR}}(\Tilde{\phi})$ in [dB] for a 1-bit DAC. Users are indicated by a star, the user angles are ${55^{\circ}, 110^{\circ}}$ in (a) and $\{  25^{\circ},  55^{\circ}, 85^{\circ}, 110^{\circ}, 135^{\circ}, 155^{\circ}\}$ in (b). }
	\label{fig:rad_zf} 
\end{figure*}

In~\cref{fig:rad_zf} the radiation pattern of the distortion, the intended signal and \gls{sdr} are depicted for the multi-user case. In~\cref{fig:rad_zf_k2} two users are present, in this case, the distortion is radiated in a number of directions, with the two user directions being the most dominant ones. \review{When the number of users increases to six in~\cref{fig:rad_zf_k6} the distortion is nearly uniformly distributed, hence receiving little to no beamforming gain. From these figures, it is clear that the distortion radiation pattern is more spatially spread out the more users are present. This is similar to the case of non-linear \gls{pa} distortion~\cite{distortion_beamformed2}. Given this spatially spread out distortion, the \gls{sdr} radiation pattern shows a clear peak in the user directions. Hence the more users are present, the less they are affected by distortion.} Note that this discussion purely focuses on the distortion, while the low resolution of \glspl{dac} can also lead to additional inter-user interference given that the low resolution leads to poorer interference cancellation. Nevertheless, when purely focusing on the distortion the most critical case arises when few users are present.

\section{Neural Networks for Precoding} \label{sec:nn_for_prec}
In this work, we consider \gls{nn}-based non-linear precoding as described by (\ref{eq:nonlin_prec}). In this section, the training procedure is outlined, followed by the \gls{nn} architecture that is selected.

 \begin{figure*}
    \centering
    \begin{tikzpicture}[scale=0.8, transform shape, node distance=1.5cm,
  thick,main node/.style={circle,draw,font=\sffamily\small\bfseries, minimum size=1cm}]


  \node[main node] (1) [minimum size=1.5cm] {$\vect{z}^{(N-1)}_{bs, 0}$};
  \node[main node] (2) [minimum size=1.5cm, below = 1cm of 1] {$\vect{z}^{(N-1)}_{bs, m}$};
  \node[main node] (4) [minimum size=1.5cm, below = 1cm of 2] {$\vect{z}^{(N-1)}_{bs, M-1}$};
  \path (2) -- (4) node [minimum size=1.175cm, font=\Large, midway, sloped] {$\dots$};
    \path (1) -- (2) node [minimum size=1.175cm, font=\Large, midway, sloped] {$\dots$};
    
  \node[above=0cm of 1] { \scriptsize$m=0$};
    \node[below=0cm of 4] { \scriptsize$m=M-1$};

  \node[main node] (5) [minimum size=1.175cm, above left = 0.05cm and 2cm  of 2] {};
  \node[main node] (6) [minimum size=1.175cm, below = 1cm of 5] {};
  \path (5) -- (6) node [thick, font=\Large, midway, sloped] {$\dots$};
    \node[above=0cm of 5] { \scriptsize$k=0$};
    \node[below=0cm of 6] { \scriptsize$k=K-1$};


  \path[every node/.style={font=\sffamily\small}]
    (1) edge[line width=0.2pt] node [sloped, left, above=0.05cm, pos=.5, fill=white, inner sep=0mm] {$(0, 0)$} (5)
    (2) edge[line width=0.2pt] node [sloped, left, above, pos=0.55, fill=white, inner sep=0mm] {} (5)
    (2) edge[line width=0.2pt] node [sloped, left, above, pos=0.35, fill=white, inner sep=0mm] {} (6)

    (4) edge[line width=0.2pt]  node [sloped, left, above, pos=0.25, fill=white, inner sep=0mm] {} (5)    
    (6) edge[line width=0.2pt]  node [sloped, left, above, pos=0.25, fill=white, inner sep=0mm] {} (4)

    (1) edge[line width=0.2pt]   node [sloped, left, above=0.05cm, pos=0.75, fill=white, inner sep=0mm] {} (6);

  \node[draw, rectangle, minimum width=4cm, minimum height=2.5cm, align=center, right = 2cm  of 2] (b1) {add Gumbel noise \\ to logits \\$g_i \sim G(0, 1)$\\$[\vect{b}]_i = [\vect{a}_{\Re, m}]_i + g_i$ \\ $\forall i \in \{0, \cdots, L-1\}$};
  \node[draw, rectangle, minimum width=5.5cm, minimum height=2.5cm, align=center, above right = -1cm and 1.3cm  of b1] (b2) {hard decision \\ 
$I=\underset{i \in\{0, \cdots, L-1\}}{\arg \max }\left\{\vect{b}\right\} \sim \mathrm{cat}\left(\vect{p}_{\Re, m}\right)$ \\
$\vect{y}_{\text {onehot }}=[0,0, \underbrace{1}_I, 0]^{\intercal}$ \\
$y_m^{\mathrm{fwd}}=\vect{y}_{\text {onehot }}^{\intercal} \mathbf{l}$};
  \node[draw, rectangle, minimum width=3cm, minimum height=2.5cm, align=center, right = 8.5cm  of b1] (b3) {compute loss \\ $J = -R_{\mathrm{sum}}(\vect{y}^{\mathrm{fwd}})$};

\path[->] (2) edge node[sloped, above] {$\vect{z}_{\mathrm{bs}, m}^{N-1} = \vect{a}_{\Re , m}$} ([yshift=01cm]b1.west);
\path[->] ([yshift=1cm]b1.east) edge node[sloped, above] {$\vect{b}$} (b2.west);
\path[->] (b2.east) edge node[sloped, above] {$y^{\mathrm{fwd}}_m$} (b3);

  \node[draw, rectangle, minimum width=5.5cm, minimum height=2.5cm, align=center, below right = -1cm and 1.3cm  of b1] (b5) {soft decision \\ $S_{i, \tau} = \frac{e^{[\vect{b}]_i/\tau}}{\sum_j e^{[\vect{b}]_j / \tau}} \forall i \in \{0, \cdots, L-1\}$ \\
  $\vect{y}_{\mathrm{soft}} = [0.1, 0.2, 0.7, 0.1]^{\intercal}$ \\ $y_m^{\mathrm{bwd}} = \vect{y}^{\intercal}_{\mathrm{soft}} \vect{l}$};

\path[<-] (2) edge node[sloped, above] {$\nabla_{\vect{a}} \vect{b}$} ([yshift=-1cm] b1.west);
\path[<-] ([yshift=-1cm]b1.east) edge node[sloped, below] {$\nabla_{\vect{b}} \vect{y}^{bwd}$} (b5.west);
\path[<-] (b5.east) edge node[sloped, below] {$\nabla_{\vect{y}^{bwd}} J$} (b3);

\path[->] (b2) edge node[right] {$\vect{b}$} (b5);

  \draw[->, dotted, bend left] (1) to[out=15, in=140] node[sloped, above, pos=0.95] {$y^{\mathrm{fwd}}_0$} (b3.north) ;

    \draw[->, dotted, bend left] (4) to[out=-15, in=-140] node[sloped, below, pos=0.95] {$y^{\mathrm{fwd}}_{M-1}$} (b3.south);

%
\def\rotangle{-10}
\node[draw, dotted, rotate=\rotangle, fill=white, minimum width=5cm, minimum height=2cm,align=center, below right = -1cm and 1.5cm of 4] (boxM){};
  \node[draw,fill=white, rotate=\rotangle, rectangle, minimum width=1cm, minimum height=0.5cm, align=center, below right = -0.25cm and 1.5cm of 4] (b1m0) {Gumbel noise};
  \node[draw, fill=white,  rotate=\rotangle, sloped, rectangle, minimum width=1cm, minimum height=0.5cm, align=center, above right = 0cm and 0cm  of b1m0] (b2m0) {hard decision};
  \node[draw, fill=white,  rotate=\rotangle, rectangle, minimum width=1cm, minimum height=0.5cm, align=center, below right = 0cm and 0cm  of b1m0] (b3m0) {soft decision};

\def\rotangle{10}
\node[draw, dotted, rotate=\rotangle, fill=white, minimum width=5cm, minimum height=2cm,align=center, below right = -2cm and 1cm of 1] (box_m0){};
  \node[draw,fill=white, rotate=\rotangle, rectangle, minimum width=1cm, minimum height=0.5cm, align=center, above right = -0.25cm and 1.5cm of 1] (b1m0) {Gumbel noise};
  \node[draw, fill=white,  rotate=\rotangle, sloped, rectangle, minimum width=1cm, minimum height=0.5cm, align=center, above right = 0cm and 0cm  of b1m0] (b2m0) {hard decision};
  \node[draw, fill=white,  rotate=\rotangle, rectangle, minimum width=1cm, minimum height=0.5cm, align=center, below right = 0cm and 0cm  of b1m0] (b3m0) {soft decision};

    \path (box_m0) -- (b1) node [minimum size=1.175cm, sloped, font=\Large, midway] {$\dots$};

    \path (boxM) -- (b1) node [minimum size=1.175cm, sloped, font=\Large, midway] {$\dots$};

\end{tikzpicture}
    \caption{Overview of the forward and backward pass during training. For clarity, only the real part is considered in this figure but the same applies for the imaginary part $\vect{a}_{\Im, m}$. The normalization step is omitted for simplicity. The vector $\vect{l}$ contains all quantization levels, hence $\vect{y}_{\mathrm{onehot}}^{\intercal} \vect{l}$ selects the desired output level. In this example, $b=2$ bits are considered, hence $L=4$. } 
    \label{fig:overview_training}
\end{figure*}
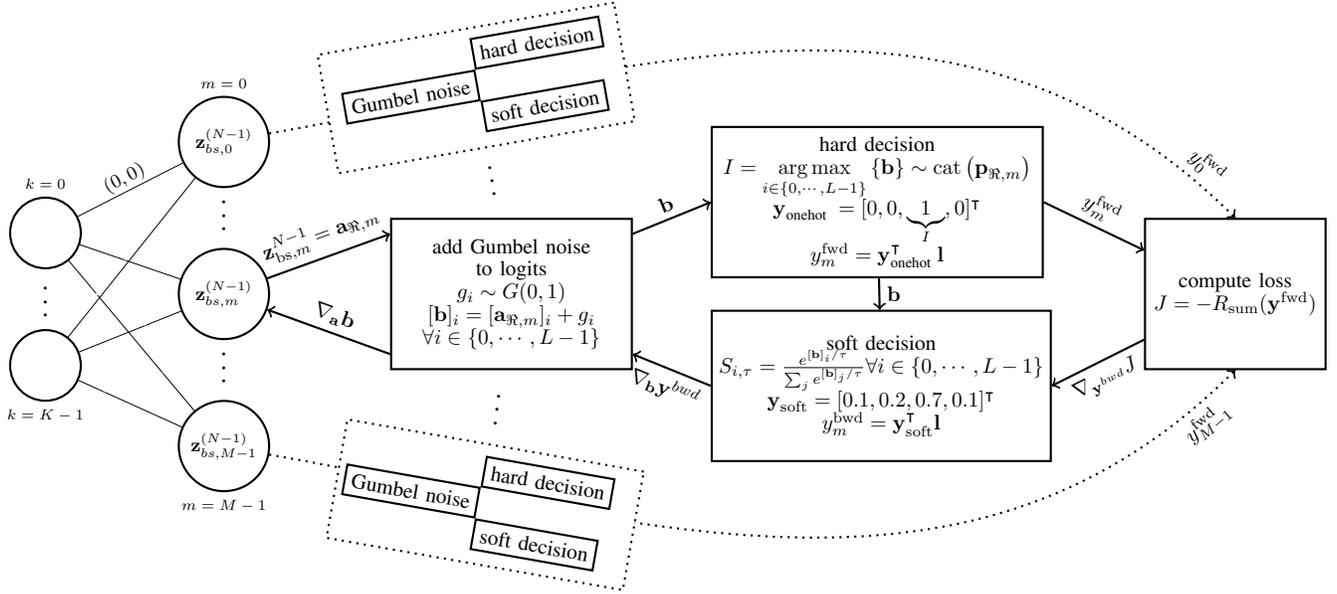

\subsection{General Training Procedure}


The optimization problem we aim to solve in order to perform quantized non-linear precoding can be formulated as
\begin{align} \label{eq:loss}
    &\max_{\vect{y}_{\mathrm{NL}}\in \mathcal{C}^{M}} \quad R_{\mathrm{sum}} \left(\vect{y}_{\mathrm{NL}}\right) \quad \mathrm{s.t.} \quad \expt{\|\vect{y}_{\mathrm{NL}}\|_2^2} \leq P_T
\end{align}
where the sum rate is computed according to (\ref{eq:sumrate}).
During training, the \gls{nn} outputs a quantized precoded vector $\vect{y}_{\mathrm{NL}}$. This precoded quantized vector can be used to compute the loss function, which is the sum rate according to (\ref{eq:sumrate})
\begin{align}
    \vect{\theta}^* = \underset{\vect{\theta}}{\arg \min } -R_{\mathrm{sum}}\left( f\left(\mat{H}, \vect{s}; \vect{\theta}\right)\right).
\end{align}
Given that the value of this sum rate directly depends on the output of the \gls{nn} (i.e., the precoded quantized vector), the gradients of this loss function, with respect to the weights of the \gls{nn} $\vect{\theta}$ can directly be computed using backpropagation. Given these gradients, the parameters of the \gls{nn} are updated using the Adam optimizer~\cite{adam}. 
Note that during training, for each channel realization, the \gls{nn} is run $N_s$ times, i.e., for $N_s$ different symbols. The average power is then normalized to $P_T$ by taking the expectation over these $N_s$ symbols according to 
\begin{align}
    \alpha &= \frac{\sqrt{P_T}}{\expt{\|\vect{y}_{\mathrm{NL}}\|_2^2}}, \quad \vect{y}_{\mathrm{norm}} = \alpha \vect{y}_{\mathrm{NL}}. \label{eq:norm}
\end{align}
Given the discrete nature of quantization, obtaining a gradient (or estimate thereof) of the loss function with respect to the weights of the \gls{nn} is non-trivial. In the next subsection, the link between quantized precoding and classification is made, followed by a method to obtain a gradient estimate, which can be used to train the \gls{nn} in an unsupervised manner.

\subsection{Unsupervised Learning of Non-Linear Precoding}
Given the discrete nature of the quantized precoding problem, the selection of the appropriate output levels can be seen as a classification problem per antenna, where the appropriate output levels need to be selected. However, the optimal output levels are unknown, hence there are no labels available to train the \gls{nn} in a supervised manner. 

\subsubsection{Intermediate Probability Mapping}
As a solution, let's consider unsupervised learning of non-linear precoding. For this, the \gls{nn} uses an intermediate mapping to probability vectors, rather than producing the correct output levels directly. 
The output of the \gls{nn} represents a probability vector per real and imaginary part of each antenna
\begin{align}
	\vect{p} &= f(\mat{H}, \vect{s}; \vect{\theta}), \quad \mathrm{with} \quad \vect{p}\mid\mat{H}, \vect{s} \\
	&= [\vect{p}_{\Re, 0}, \vect{p}_{\Im, 0}, \cdots , \vect{p}_{\Re, M-1}, \vect{p}_{\Im, M-1}]^{\intercal}
\end{align}
where the intermediate mapping is $f(\cdot, \cdot): \mathbb{C}^{M\times K} \times \mathbb{C}^{K} \mapsto [0, 1]^{2M|\mathcal{L}|}$ and $\vect{p} \in [0, 1]^{2M|\mathcal{L}|}$ represents the concatenation of the probability vectors for the real and imaginary parts of each antenna, where $\vect{p}_{\Re, m}, \vect{p}_{\Im, m} \in [0, 1]^{|\mathcal{L}|}$ represent the probability vectors for the real and imaginary parts of the \gls{dac} output at antenna $m$. These probability vectors are conditioned on the channel matrix and symbol vector. However, this conditioning is often omitted for notational convenience. At inference time, the true output levels of \gls{dac} $m$ can be found as the output levels $l_i, l_j \in \mathcal{L}$ that are assigned the highest probability value. This gives $y_m = l_i + l_j \jmath$, with
 \begin{equation}\label{eq:inference}
\begin{aligned}
    i &= \argmax_{i\in  \{0, \cdots , L-1 \}} \vect{p}_{\Re, m} , \quad
    j = \argmax_{j \in  \{0, \cdots , L-1 \}}  \vect{p}_{\Im, m}. 
\end{aligned}
\end{equation}

Each of the obtained probability vectors is a result of a softmax activation. The elements of these vectors can be interpreted as the parameters of a categorical distribution conditioned on $\mat{H}$ and $\vect{s}$, where each category (output level in our case) is assigned a certain probability of being the correct output level.  The logits/the raw outputs of the \gls{nn} (i.e., the output before applying the softmax activation) for the real and imaginary parts of antenna $m$ are defined as $\vect{a}_{\Re, m}$, $\vect{a}_{\Im, m}$ respectively. The softmax of vector $\vect{a}_{\Re, m}$ is computed as\footnote{For the remainder of this discussion we will only refer to the real part of antenna $m$, $\vect{a}_{\Re, m}$, however the same applies to the imaginary part $\vect{a}_{\Im, m}$.} 
\begin{align}
	\left[\vect{p}_{\Re, m}\right]_i = \left[\mathrm{softmax}( \vect{a}_{\Re, m})\right]_i = \frac{e^{[ \vect{a}_{\Re, m}]_i}}{\sum_j e^{[ \vect{a}_{\Re, m}]_j}} \label{eq:softmax}
\end{align} 
where $\left[\mathrm{softmax}( \vect{ a}_{\Re, m})\right]_i$ is the i-th element of the softmax-transformed vector $\vect{p}_{\Re, m}$. This scales all values of $\vect{a}_{\Re, m}$ between zero and one and ensures that the summation of all elements of the vector equals one, hence the output values of the softmax function can be interpreted as probabilities.





\subsubsection{Overcoming Non-Differentiability}
If we know the correct labels for $\vect{p}_{\Re, m}, \vect{p}_{\Im, m}$ we can learn in a supervised manner by minimizing the cross entropy loss between the output of the softmax layers and the labels. Given that these labels are expensive to obtain, unsupervised training is used. We interpret the output of the softmax layer as the probability each output level has to be the 'best' value. To select the best output labels we take the argmax of these probability vectors to select the maximum likelihood output levels, as depicted in \eqref{eq:inference}.
The problem here is that the argmax function is non-differentiable. A common workaround is to use the argmax function in the forward pass of the \gls{nn} and estimate the gradients using a softmax function in the backward pass. \review{However, empirically, we found this approach to give poor results, as it often leads to greedy, sub-optimal solutions~\cite{gumbel_review}. Motivated by the exploration-exploitation dilemma, we aim to introduce more exploration by treating the output of the \gls{nn} as the parameters of a categorical distribution, from which samples are drawn rather than selecting the maximum value via argmax.  This sampling introduces a degree of exploration, enabling the \gls{nn} to observe a wider variety of (potentially negative) examples during training. This enables the model to escape sub-optimal local minima. By contrast, relying on the deterministic argmax function emphasizes exploitation, which risks converging to greedy, suboptimal solutions~\cite{gumbel_review}. }



To train the \gls{nn} in this way a sampling scheme of which the gradients can be estimated is required. We use a relaxed reparameterization gradient estimator known as the straight-through Gumbel-softmax estimator~\cite{gumbel_review, gumbel_softmax}. To build up to this we discuss three variants: \textit{(1)} the Gumbel-max estimator, \textit{(2)} the Gumbel-softmax estimator and \textit{(3)} the straight-through Gumbel-softmax estimator. 
First, we consider a reparameterization gradient estimator \textit{(1)}. This estimator reparameterizes the sampling of a random variable by using a sample from a standard distribution combined with a deterministic transformation. In our case, this is referred to as the Gumbel-max trick, where a sample from the categorical distribution $I \sim \mathrm{cat}(\vect{p}_{\Re, m})$ is obtained by sampling the standard Gumbel distribution and then transforming it. More concretely, sampling the categorial distribution is equivalent to adding samples from the standard Gumbel distribution to the logits and selecting the index with the highest value~\cite{gumbel_review}
\begin{align}
    I = \argmax_{i\in  \{0, \cdots , L-1 \}}\{[\vect{a}_{\Re, m}]_i + g_i\} \sim \mathrm{cat}(\vect{p}_{\Re, m})
\end{align}
where $g_i\sim\mathrm{Gumbel}(0, 1)$ are samples from the standard Gumbel distribution and $[\vect{a}_{\Re , m}]_i$ are unnormalized probabilities (i.e., the input to the softmax function as in (\ref{eq:softmax})). However, for the Gumbel-max trick, the deterministic transformation is non-differentiable due to the argmax function. Therefore, a relaxed gradient estimator \textit{(2)} is often used which replaces the argmax by a softmax function (typically with an added temperature variable $\tau$). These soft samples are defined as
\begin{align}
	S_{i,\tau} = \frac{e^{([\vect{a}_{\Re, m}]_i + g_i)/\tau}}{\sum_j e^{([\vect{a}_{\Re, m}]_j + g_j)/\tau}}  \quad \mathrm{for} \ i\in \{0, \cdots , L-1 \}.
\end{align}
This approximates the (typically discrete) sample of the categorical distribution by a continuous random variable, which allows for the computation of gradients. This is known as the Gumbel-softmax trick. In the limit where $\tau \to 0$ this converges to a discrete (one-hot encoded) sample $\mathbbm{1}_i^{|\mathcal{L}|}$. For some applications, this relaxation is acceptable, however, given that the aim of this work is quantization, a continuous relaxation would undo the desired quantization. To overcome this, we use the straight-through Gumbel-softmax estimator \textit{(3)}. This estimator produces discrete samples in the forward pass and uses the gradients from the Gumbel-softmax trick in the backward pass. 
This can effectively be seen as using the Gumbel-max trick in the forward pass, and the Gumbel-softmax trick in the backward pass. An overview of this training procedure can be seen in~\cref{fig:overview_training}. Note that, the straight-through Gumbel-softmax estimator is only used during training. At inference time, the argmax function is used to select the output levels that are assigned the highest probability according to~\eqref{eq:inference}, to obtain the precoded quantized vector $\vect{y}$.

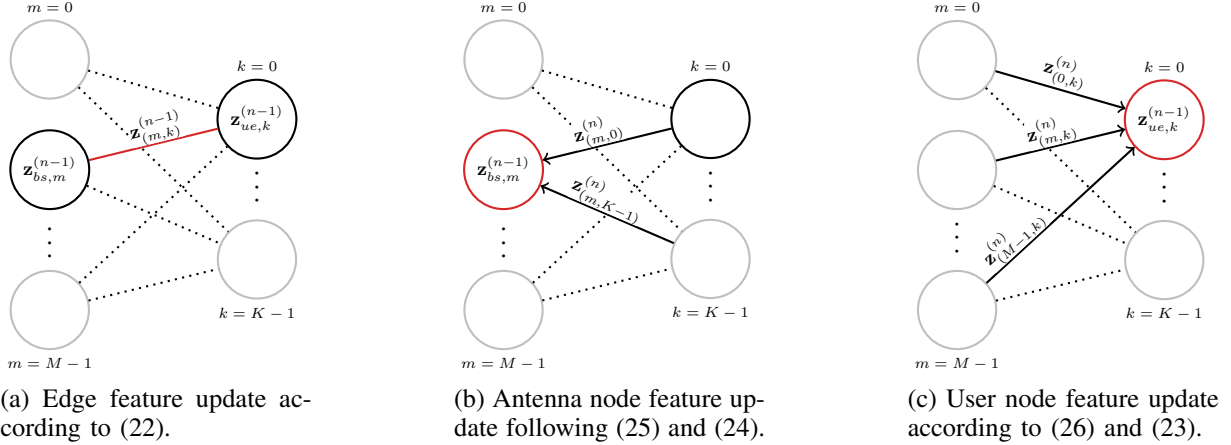
\begin{figure*}[t]
    \centering
    \begingroup
    \tikzset{every picture/.style={scale=0.8}}%
    \hspace{0.05 \linewidth}
    \subfloat[Edge feature update according to (\ref{eq:gnn1}).]{\label{fig:overview_a}\begin{tikzpicture}[transform shape, node distance=1.5cm,
  thick,main node/.style={circle,draw,font=\sffamily\small\bfseries, minimum size=1cm}]

 \def\nodesize{1.3}
\definecolor{myred}{HTML}{d62728}
\definecolor{myblue}{HTML}{1f77b4}
\definecolor{myolive}{HTML}{bcbd22}
\definecolor{mypurple}{HTML}{9467bd}
\definecolor{myorange}{HTML}{ff7f0e}
\definecolor{mypink}{HTML}{e377c2}
\definecolor{mygreen}{HTML}{2ca02c}
\definecolor{mybrown}{HTML}{8c564b}
\definecolor{mygray}{HTML}{7f7f7f}
\definecolor{mycyan}{HTML}{17becf}

 \node[main node, draw=lightgray] (1) [minimum size=\nodesize cm] {};
  \node[main node, ] (2) [minimum size=\nodesize cm, below = 0.5cm of 1] {$\vect{z}^{(n-1)}_{bs, m}$};
  \node[main node, draw=lightgray] (3) [minimum size=\nodesize cm, below = 1cm of 2] {};
  \path (2) -- (3) node [minimum size=\nodesize cm, font=\Large, midway, sloped] {$\dots$};
  \node[above=0cm of 1] { \scriptsize$m=0$};
    \node[below=0cm of 3] { \scriptsize$m=M-1$};
  \node[main node] (4) [minimum size=\nodesize cm, below right = 0.05cm and 2.5cm  of 1] {$\vect{z}^{(n-1)}_{ue, k}$};
  \node[main node, draw=lightgray] (5) [minimum size=\nodesize cm, below = 1cm of 4] {};
  \path (4) -- (5) node [thick, font=\Large, midway, sloped] {$\dots$};
      \node[above=0cm of 4] { \scriptsize$k=0$};
    \node[below=0cm of 5] { \scriptsize$k=K-1$};


  \path[every node/.style={font=\sffamily\small}]
    (1) edge [dotted] node [sloped, left, above=0.05cm, pos=.5, fill=white, inner sep=0mm] {} (4)
    (1) edge [dotted] node [sloped, left, above=0.05cm, pos=.5, fill=white, inner sep=0mm] {} (5)

    (2) edge[draw=myred] node [sloped, left, above, pos=0.55, fill=white, inner sep=0mm] {$\vect{z}^{(n-1)}_{(m,k)}$} (4)
    (2) edge[dotted] node [sloped, left, above, pos=0.35, fill=white, inner sep=0mm] {} (5)

    (3) edge[dotted]  node [sloped, left, above, pos=0.25, fill=white, inner sep=0mm] {} (4)    
    (3) edge[dotted]  node [sloped, left, above, pos=0.25, fill=white, inner sep=0mm] {} (5);

\end{tikzpicture}}\hfill
    \subfloat[Antenna node feature update following (\ref{eq:gnn4}) and (\ref{eq:gnn3}).]{\label{fig:overview_b}\begin{tikzpicture}[transform shape, node distance=1.5cm,
  thick,main node/.style={circle,draw,font=\sffamily\small\bfseries, minimum size=1cm}]

 \def\nodesize{1.3}
\definecolor{myred}{HTML}{d62728}
\definecolor{myblue}{HTML}{1f77b4}
\definecolor{myolive}{HTML}{bcbd22}
\definecolor{mypurple}{HTML}{9467bd}
\definecolor{myorange}{HTML}{ff7f0e}
\definecolor{mypink}{HTML}{e377c2}
\definecolor{mygreen}{HTML}{2ca02c}
\definecolor{mybrown}{HTML}{8c564b}
\definecolor{mygray}{HTML}{7f7f7f}
\definecolor{mycyan}{HTML}{17becf}

 \node[main node, draw=lightgray] (1) [minimum size=\nodesize cm] {};
  \node[main node, draw=myred] (2) [minimum size=\nodesize cm, below = 0.5cm of 1] {$\vect{z}^{(n-1)}_{bs, m}$};
  \node[main node, draw=lightgray] (3) [minimum size=\nodesize cm, below = 1cm of 2] {};
  \path (2) -- (3) node [minimum size=\nodesize cm, font=\Large, midway, sloped] {$\dots$};
  \node[above=0cm of 1] { \scriptsize$m=0$};
    \node[below=0cm of 3] { \scriptsize$m=M-1$};
  \node[main node] (4) [minimum size=\nodesize cm, below right = 0.05cm and 2.5cm  of 1] {};
  \node[main node, draw=lightgray] (5) [minimum size=\nodesize cm, below = 1cm of 4] {};
  \path (4) -- (5) node [thick, font=\Large, midway, sloped] {$\dots$};
      \node[above=0cm of 4] { \scriptsize$k=0$};
    \node[below=0cm of 5] { \scriptsize$k=K-1$};


  \path[every node/.style={font=\sffamily\small}]
    (1) edge [dotted] node [sloped, left, above=0.05cm, pos=.5, fill=white, inner sep=0mm] {} (4)
    (1) edge [dotted] node [sloped, left, above=0.05cm, pos=.5, fill=white, inner sep=0mm] {} (5)
    (3) edge[dotted]  node [sloped, left, above, pos=0.25, fill=white, inner sep=0mm] {} (4)    
    (3) edge[dotted]  node [sloped, left, above, pos=0.25, fill=white, inner sep=0mm] {} (5)
    
    (2) edge[<-] node [sloped, left, above, pos=0.46, fill=white, inner sep=0mm] {$\vect{z}^{(n)}_{(m,0)}$} (4)
    (2) edge[<-] node [sloped, left, above, pos=0.46, fill=white, inner sep=0mm] {$\vect{z}^{(n)}_{(m,K-1)}$} (5);

\end{tikzpicture}}\hfill
     \subfloat[User node feature update according to (\ref{eq:gnn5}) and (\ref{eq:gnn2}).]{\label{fig:overview_c}\begin{tikzpicture}[transform shape, node distance=1.5cm,
  thick,main node/.style={circle,draw,font=\sffamily\small\bfseries, minimum size=1cm}]

 \def\nodesize{1.3}
\definecolor{myred}{HTML}{d62728}
\definecolor{myblue}{HTML}{1f77b4}
\definecolor{myolive}{HTML}{bcbd22}
\definecolor{mypurple}{HTML}{9467bd}
\definecolor{myorange}{HTML}{ff7f0e}
\definecolor{mypink}{HTML}{e377c2}
\definecolor{mygreen}{HTML}{2ca02c}
\definecolor{mybrown}{HTML}{8c564b}
\definecolor{mygray}{HTML}{7f7f7f}
\definecolor{mycyan}{HTML}{17becf}

 \node[main node, draw=lightgray] (1) [minimum size=\nodesize cm] {};
  \node[main node, draw=lightgray] (2) [minimum size=\nodesize cm, below = 0.5cm of 1] {};
  \node[main node, draw=lightgray] (3) [minimum size=\nodesize cm, below = 1cm of 2] {};
  \path (2) -- (3) node [minimum size=\nodesize cm, font=\Large, midway, sloped] {$\dots$};
  \node[above=0cm of 1] { \scriptsize$m=0$};
    \node[below=0cm of 3] { \scriptsize$m=M-1$};
  \node[main node, draw=myred] (4) [minimum size=\nodesize cm, below right = 0.05cm and 2.5cm  of 1] {$\vect{z}^{(n-1)}_{ue, k}$};
  \node[main node, draw=lightgray] (5) [minimum size=\nodesize cm, below = 1cm of 4] {};
  \path (4) -- (5) node [thick, font=\Large, midway, sloped] {$\dots$};
      \node[above=0cm of 4] { \scriptsize$k=0$};
    \node[below=0cm of 5] { \scriptsize$k=K-1$};


  \path[every node/.style={font=\sffamily\small}]
    (1) edge [->] node [sloped, left, above=0.05cm, pos=.5, fill=white, inner sep=0mm] {$\vect{z}^{(n)}_{(0,k)}$} (4)
    (1) edge [dotted] node [sloped, left, above=0.05cm, pos=.5, fill=white, inner sep=0mm] {} (5)
    (3) edge[->]  node [sloped, left, above, pos=0.25, fill=white, inner sep=0mm] {$\vect{z}^{(n)}_{(M-1,k)}$} (4)    
    (3) edge[dotted]  node [sloped, left, above, pos=0.25, fill=white, inner sep=0mm] {} (5)
    
    (2) edge[->] node [sloped, left, above, pos=0.46, fill=white, inner sep=0mm] {$\vect{z}^{(n)}_{(m,k)}$} (4)
    (2) edge[dotted] node [sloped, left, above, pos=0.46, fill=white, inner sep=0mm] {} (5);

\end{tikzpicture}}
    \hspace{0.05 \linewidth}

     \endgroup
	\caption{\Gls{gnn} overview: antenna nodes are on the left, user nodes on the right. 
    Red indicates the element being updated, black the elements used to perform that update. The previous value of the element, before the update, is also used during the update. }
	\label{fig:overview_gnn} 
\end{figure*}

\subsection{Neural Network Architecture}
\review{
 In this section, we discuss which type of \gls{nn} architecture is adequate for non-linear precoding. Fully connected neural networks, with 1 hidden layer and sufficient neurons, can approximate any continuous non-linear function with arbitrary accuracy~\cite{universalapprox}. This makes them theoretically capable of learning the desired function mapping concerning the problem considered in our work, i.e., a continuous mapping from channel matrix and symbol vector to probability vectors over the possible output levels of the \glspl{dac}. Unfortunately, the universal approximation theorem is not constructive, i.e., it does not indicate how to choose the network architecture or the number of neurons required to achieve such accuracy. Given the universal function approximation theorem, the hypothesis space of a sufficiently large fully connected \gls{nn}, that is the set of all possible functions the fully connected \gls{nn} can represent, is theoretically infinite. While this expressive power is appealing, it introduces practical challenges. A larger hypothesis space increases the number of trainable parameters, making the network more difficult to train, data-hungry, and computationally expensive during inference. As such, in practice, it is advantageous to reduce the size of this hypothesis space in an informed manner. For a formal discussion of this mechanism, based on statistical learning theory, we refer to~\cite{gnn_feys, cnn_gnn, understanding_gnn_precoding}. However, the main insight is as follows: by using certain \gls{nn} architectures we restrict the \gls{nn} to only learn certain groups of functions. As such, the hypothesis space of the \gls{nn} is greatly reduced. This leads to more scalable \glspl{nn} with fewer trainable parameters, which are easier to train and have lower computational complexity. This comes at the cost of a less expressive model, as the model can only learn the limited group of functions. The main idea lies in selecting the right \gls{nn} architecture, which can learn a group of functions, which is known to cover the desired function. In this way, the reduction in expressivity does not affect the performance, as only functions outside of this group are ruled out. Next, we explain how this mechanism can be applied to our problem by limiting the \gls{nn} to only learn permutation equivariant/invariant functions.
 
 For non-linear precoding, given the physical underlying system, the following equivariance and invariance properties need to be maintained. \textit{i)} If the user order is permuted with permutation matrix $\mat{\Pi} \in \{0, 1\}^{K\times K}$, the output of the precoding function should remain unchanged, hence $f(\mat{H \Pi}, \mat{\Pi}\vect{s}) = \vect{y}_{\mathrm{NL}}$. The learned function should thus be permutation invariant with respect to the symbol vector and columns of the channel matrix. \textit{ii)} If the antenna ordering is permuted with permutation matrix $\mat{\Pi} \in \{0, 1\}^{M\times M}$, the output of the precoding function should be permuted accordingly $f(\mat{\Pi H}, \vect{s}) = \mat{\Pi} \vect{y}_{\mathrm{NL}}$. The learned function should thus be permutation equivariant with respect to the rows of the channel matrix. When a \gls{nn} imposes these properties on the learned functions, its hypothesis space is reduced as it can no longer learn all existing functions, but only the subset that respects these properties. Furthermore, this can only positively affect the learning performance, as the optimal function (i.e., a function that is permutation equivariant with respect to the antenna ordering, and permutation invariant with respect to the user ordering), is still covered by this restricted hypothesis space.}

Next, a \gls{gnn} architecture is described which respects these properties. The proposed \gls{gnn} is based on generalized message passing~\cite{graph_rep_learning, inductivebias}. 
The \gls{gnn} learns the mapping from channel matrix and symbol vector to a precoded and quantized vector over a graph. The graph in question has a node for each transmit antenna at the \gls{bs}, a node for each user and an edge between each antenna and user as can be seen in~\cref{fig:overview_a}. The graph is defined as $\mathcal{G} = (\mathcal{V}, \mathcal{E})$, where $\mathcal{V} = \mathcal{V}_M \cup \mathcal{V}_K$ is the set of nodes, with $\mathcal{V}_M$ the antenna nodes and $\mathcal{V}_K$ the user nodes, and $\mathcal{E}$ the set of edges, which contains all edges between the antenna and user nodes i.e., $(m, k) \in \mathcal{E}  \ \forall m \in \mathcal{V}_M, k \in \mathcal{V}_K$. The graph is undirected, meaning that $(m, k) \in \mathcal{E} \leftrightarrow (k, m) \in \mathcal{E}$, as is the physical channel between each user and antenna. A general \gls{gnn} performs a number of message-passing iterations. At iteration/layer $n$, a hidden representation is updated for each edge, antenna node, and user node, denoted by $\vect{z}^{(n)}_{(m,k)}$, $\vect{z}^{(n)}_{bs, m}$ and $\vect{z}^{(n)}_{ue, k} \in \mathbb{R}^{d_n}$ respectively. These representations are updated in layer $n$ according to
\begin{align}
	\vect{z}^{(n)}_{(m,k)} &= \sigma\left(\mat{W}_{\mathrm{edge}}^{(n)} \vect{z}_{(m,k)}^{(n-1)}+ \mat{W}_{\mathrm{bs}}^{(n)} \vect{z}_{bs, m}^{(n-1)}+    \mat{W}_{\mathrm{ue}}^{(n)} \vect{z}_{ue, k}^{(n-1)} \right)\label{eq:gnn1}\\ 
	\vect{m}_{\mathcal{N}(k)} &= \frac{1}{|\mathcal{N}(k)|} \sum_{m^{\prime}\in\mathcal{N}(k)} \vect{z}_{(m^{\prime}, k)}^{(n)}\label{eq:gnn2}\\
	\vect{m}_{\mathcal{N}(m)} &= \frac{1}{|\mathcal{N}(m)|} \sum_{k^{\prime}\in\mathcal{N}(m)} \vect{z}_{(m, k^{\prime})}^{(n)}\label{eq:gnn3}\\
	\vect{z}_{bs, m}^{(n)} &= \sigma\left(\mat{W}_{\mathrm{self, bs}}^{(n)} \vect{z}_{bs, m}^{(n-1)} + \mat{W}_{\mathrm{neigh, bs}}^{(n)} \vect{m}_{\mathcal{N}(m)}\right) \label{eq:gnn4} \\
	\vect{z}_{ue, k}^{(n)} &= \sigma\left(\mat{W}_{\mathrm{self, ue}}^{(n)} \vect{z}_{ue, k}^{(n-1)} + \mat{W}_{\mathrm{neigh, ue}}^{(n)} \vect{m}_{\mathcal{N}(k)}\right).\label{eq:gnn5}
\end{align}
 Here $\sigma(\cdot)$ is a non-linear activation and the matrices $\mat{W}_{\mathrm{edge}}^{(n)}, \mat{W}_{\mathrm{bs}}^{(n)}, \mat{W}_{\mathrm{ue}}^{(n)}, \mat{W}_{\mathrm{self, bs}}^{(n)}, \mat{W}_{\mathrm{self, ue}}^{(n)} \in \mathbb{R}^{d_{n} \times d_{n-1}}$ and $\mat{W}_{\mathrm{neigh, bs}}^{(n)}$, $\mat{W}_{\mathrm{neigh, ue}}^{(n)} \in \mathbb{R}^{d_{n} \times d_{n}}$ are learned. The dimensions of these matrices ($d_{n-1}$, $d_n$) determine the number of features in layer $n-1$ and $n$ respectively. Note that $m$ denotes the antenna index, $k$ the user index and $(m,k)$ the edge between antenna $m$ and user $k$. First, in (\ref{eq:gnn1}) the edge features are updated according to the edge, antenna and user features from the previous layer. Second, in (\ref{eq:gnn2}) message passing is performed from the neighbors of each user node to generate the user messages, in (\ref{eq:gnn3}) message passing is performed from the neighbors of each antenna node to generate the antenna messages. Next, in (\ref{eq:gnn4}) the antenna features are updated according to the antenna messages and the antenna features from the previous layer. Finally, in (\ref{eq:gnn5}) the user features are updated according to the user features from the previous layer and the user messages. \Cref{fig:overview_gnn} provides an overview of these computations. Note that these operations are repeated for a total of $N$ layers: one input layer, one output layer and $N_h = N-2$ hidden layers.


The inputs for edge $(m,k)$, antenna node $m$ and user node $k$ in layer one of the \gls{gnn} are defined in the following way 
\begin{align}
    \mathbf{z}_{(m, k)}^{(0)} &= \begin{bmatrix} \Re \{[\mathbf{H}]_{m,k}\} , \Im \{[\mathbf{H}]_{m,k} \}\end{bmatrix}^{\intercal}\\
     \mathbf{z}_{bs, m}^{(0)} &= \begin{bmatrix} 0, 0\end{bmatrix}^{\intercal}\\
      \mathbf{z}_{ue, k}^{(0)} &= \begin{bmatrix} \Re \{s_k\} , \Im \{s_k \}\end{bmatrix}^{\intercal}.
\end{align}
The outputs of the \gls{gnn} are the precoded symbols, which are defined as the antenna node features in the final layer 
\begin{align}
        \mathbf{z}_{bs, m}^{(N-1)} &= \begin{bmatrix} \Re\{y_m\} , \Im\{y_m\}\end{bmatrix}^{\intercal}.
\end{align}

Note, however, that the actual outputs of the \gls{gnn} are probability vectors as discussed in the previous section
$\mathbf{z}_{bs, m}^{(N-1)} = \begin{bmatrix} \vect{p}_{\Re, m} , \vect{p}_{\Im, m}\end{bmatrix}^{\intercal}$.
At inference time, these probability vectors are mapped to the precoded symbols according to \eqref{eq:inference}.


\section{Complexity Analysis}
\label{sec:complexity}

In this section, the complexity of the proposed GNN is computed in terms of real \gls{flops}. 

\subsection{Complexity of the GNN}
The forward pass of the \gls{gnn} is described in equations (\ref{eq:gnn1}), (\ref{eq:gnn2}), (\ref{eq:gnn3}), (\ref{eq:gnn4}) and (\ref{eq:gnn5}). To compute the complexity, a distinction is made between the input layer, the four hidden layers ($N_h=4$) and the output layer, as the dimensions of the inputs, the features, and the outputs are different in each of these layers. In the input layer $d_0 = 2$, in the hidden layers $d_n = d_h = 128 \ \forall n \in \{1, \cdots, N-2\}$ and in the output layer $d_{N-1} = 2^{b+1}$. Additionally, (\ref{eq:gnn1}) has to be computed for each edge ($MK$ times), (\ref{eq:gnn2}) and (\ref{eq:gnn5}) for each user node ($K$ times) and (\ref{eq:gnn3}) and (\ref{eq:gnn4}) for each antenna node ($M$ times).

For the input layer, the number of real floating-point multiplications is
\begin{align}
\begin{split}
    \mathcal{O}_{\mathrm{in}}^{(\mathrm{mul})} = &\underbrace{6MKd_h}_{(\ref{eq:gnn1})} + \underbrace{Kd_h}_{(\ref{eq:gnn2})} + \underbrace{Md_h}_{(\ref{eq:gnn3}} \\&+ \underbrace{2Md_h + Md_h^2}_{(\ref{eq:gnn4})} + \underbrace{2Kd_h + Kd_h^2}_{(\ref{eq:gnn5})}
    \end{split}
\end{align}
and the number of real floating-point additions is
\begin{align}
\begin{split}
    \mathcal{O}_{\mathrm{in}}^{(\mathrm{add})} = &\underbrace{5MKd_h}_{(\ref{eq:gnn1})} + \underbrace{K(M-1)d_h}_{(\ref{eq:gnn2})} + \underbrace{M(K-1)d_h}_{(\ref{eq:gnn3})} \\ &+ \underbrace{Md_h + Md_h^2}_{(\ref{eq:gnn4})} + \underbrace{Kd_h + Kd_h^2}_{(\ref{eq:gnn5})},
\end{split}
\end{align}
where the underbraces indicate the equations that require the indicated computations.
One hidden layer requires 
\begin{align}
    \mathcal{O}_{\mathrm{h}}^{(\mathrm{mul})} = \underbrace{3MKd_h^2}_{(\ref{eq:gnn1})} + \underbrace{Kd_h}_{(\ref{eq:gnn2})} + \underbrace{Md_h}_{(\ref{eq:gnn3})} + \underbrace{2Md_h^2}_{(\ref{eq:gnn4})} + \underbrace{2Kd_h^2}_{(\ref{eq:gnn5})}
\end{align}
real floating-point multiplications and 
\begin{align}
\begin{split}
    \mathcal{O}_{\mathrm{h}}^{(\mathrm{add})} =&\underbrace{3MKd_h^2 - MKd_h}_{(\ref{eq:gnn1})} + \underbrace{K(M-1)d_h}_{(\ref{eq:gnn2})} + \underbrace{M(K-1)d_h}_{(\ref{eq:gnn3})} \\ &+ \underbrace{2Md_h^2 - Md_h}_{(\ref{eq:gnn4})} + \underbrace{2Kd_h^2 - Kd_h}_{(\ref{eq:gnn5})}
\end{split}
\end{align}
real floating-point additions.
For the output layer, only the antenna node features are computed as they are the final output of the \gls{gnn}. This requires
\begin{align}
    \mathcal{O}_{\mathrm{out}}^{(\mathrm{mul})} = \underbrace{6MKd_h2^b}_{(\ref{eq:gnn1})} + \underbrace{2M2^b}_{(\ref{eq:gnn3})} + \underbrace{4M2^{2b} +2Md_h2^b}_{(\ref{eq:gnn4})}
\end{align}
real floating-point multiplications and
\begin{align}
\begin{split}
    \mathcal{O}_{\mathrm{out}}^{(\mathrm{add})} = &\underbrace{6MKd_h2^b - 2MK2^b}_{(\ref{eq:gnn1})} \\&+ \underbrace{2M(K-1) 2^b}_{(\ref{eq:gnn3})} + \underbrace{4M 2^{2b} + 2Md_h2^b - 2M2^b}_{(\ref{eq:gnn4})}
    \end{split}
\end{align}
real floating-point additions. In total, the \gls{gnn} with $N_h=4$ requires the following number of floating-point multiplications
\begin{align}\label{eq:compl_mul}
   \mathcal{O}_{\mathrm{GNN}}^{(\mathrm{mul})} &=  \mathcal{O}_{\mathrm{in}}^{(\mathrm{mul})} + N_h \mathcal{O}_{\mathrm{h}}^{(\mathrm{mul})} + \mathcal{O}_{\mathrm{out}}^{(\mathrm{mul})} 
\end{align}

and the following number of floating-point additions
\begin{align}\label{eq:compl_add}
   \mathcal{O}_{\mathrm{GNN}}^{(\mathrm{add})} &=  \mathcal{O}_{\mathrm{in}}^{(\mathrm{add})} + N_h \mathcal{O}_{\mathrm{h}}^{(\mathrm{add})} + \mathcal{O}_{\mathrm{out}}^{(\mathrm{add})} 
\end{align}
This scales as $\mathcal{O}(d_h^2MK + 2^bMKd_h + 2^{2b}M) $ multiplications and $\mathcal{O}(d_h^2MK + 2^bMKd_h + 2^{2b}M)$ additions.

\section{Simulation Results}\label{sec:results}

\subsection{Training Parameters and Simulation Setup}

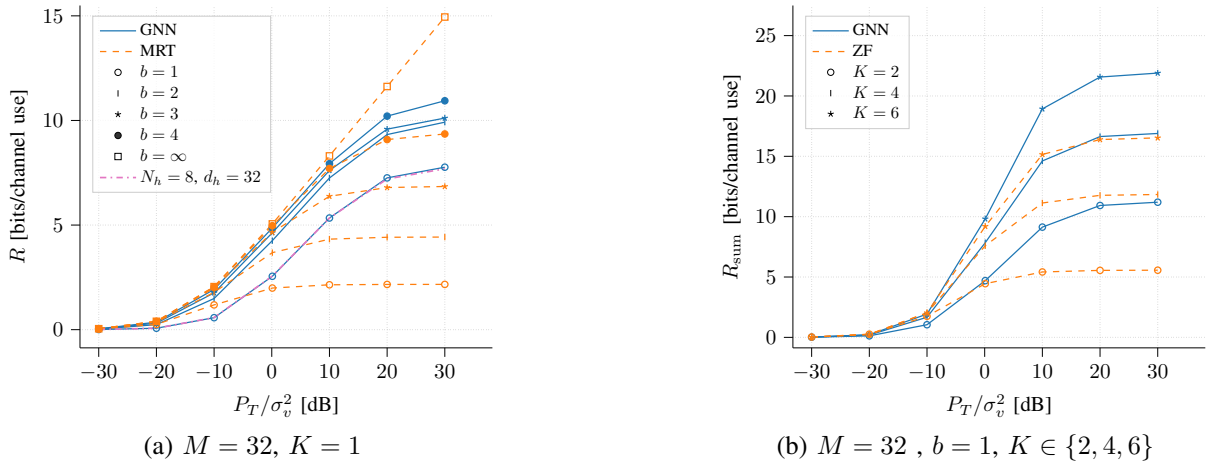
\begin{figure*}[ht!]
    \centering
    \subfloat[$M=32$, $K=1$]{\label{fig:rsum_k1_b1}
\begin{tikzpicture}[scale=0.8]

\definecolor{crimson2143940}{RGB}{214,39,40}
\definecolor{darkgray176}{RGB}{176,176,176}
\definecolor{darkorange25512714}{RGB}{255,127,14}
\definecolor{forestgreen4416044}{RGB}{44,160,44}
\definecolor{lightgray204}{RGB}{204,204,204}
\definecolor{steelblue31119180}{RGB}{31,119,180}
\definecolor{green01270}{RGB}{0,127,0}

\definecolor{myred}{HTML}{d62728}
\definecolor{myblue}{HTML}{1f77b4}
\definecolor{myolive}{HTML}{bcbd22}
\definecolor{mypurple}{HTML}{9467bd}
\definecolor{myorange}{HTML}{ff7f0e}
\definecolor{mypink}{HTML}{e377c2}
\definecolor{mygreen}{HTML}{2ca02c}
\definecolor{mybrown}{HTML}{8c564b}
\definecolor{mygray}{HTML}{7f7f7f}
\definecolor{mycyan}{HTML}{17becf}

\begin{axis}[
axis lines* = {left},
legend cell align={left},
legend style={nodes={scale=0.8, transform shape},
  fill opacity=0.8,
  draw opacity=1,
  text opacity=1,
  at={(0.03,0.97)},
  anchor=north west,
  draw=lightgray204
},
tick align=outside,
tick pos=left,
title={},
   x grid style={white!69.01960784313725!black},
        grid,
        grid style={help lines,color=gray!50, densely dotted},
        xlabel={$P_T/\sigma_v^2$ [dB]},
xmin=-33.25, xmax=38.25,
xtick style={color=black},
y grid style={darkgray176},
ylabel={$R$ [bits/channel use]},
ymin=-0.869273127070603, ymax=15.4946536076367,
ytick style={color=black}
]

\addlegendimage{semithick, myblue}
\addlegendentry{GNN}

\addlegendimage{semithick, dashed, myorange}
\addlegendentry{MRT}

\addlegendimage{only marks, black, mark=o, mark size=1.5, mark options={solid}, }
\addlegendentry{$b=1$}

\addlegendimage{only marks, black, mark=|, mark size=1.5, mark options={solid}, }
\addlegendentry{$b=2$}

\addlegendimage{only marks, black, mark=star, mark size=1.5, mark options={solid}, }
\addlegendentry{$b=3$}

\addlegendimage{only marks, black, mark=*, mark size=1.5, mark options={solid}, }
\addlegendentry{$b=4$}

\addlegendimage{only marks, black, mark=square, mark size=1.5, mark options={solid}, }
\addlegendentry{$b=\infty$}

\addlegendimage{thick, dashdotted, mypink}
\addlegendentry{$N_h=8$, $d_h=32$}

\addplot [semithick, myblue, mark=o, mark size=1.5, mark options={solid}, mark repeat={1}]
table {%
-30 0.00703622214241695
-20 0.0688368199192074
-10 0.572434705972911
0.1 2.5520370966241
10 5.33707171225104
20 7.25474323296877
30 7.7659708214884
};
\addplot [semithick, myblue, mark=|, mark size=1.5, mark options={solid}, mark repeat={1}]
table {%
-30 0.0260358496028232
-20 0.241052945027971
-10 1.48923767610045
0.1 4.25639591570041
10 7.25062726971131
20 9.3257330108142
30 9.91737847384212
};

\addplot [semithick, myblue, mark=star, mark size=1.5, mark options={solid}, mark repeat={1}]
table {%
-30 0.0344571548065754
-20 0.311571347654304
-10 1.76016516826574
0.1 4.63363529215551
10 7.61099120725961
20 9.58462843858504
30 10.1133234994399
};

\addplot [semithick, myblue, mark=*, mark size=1.5, mark options={solid}, mark repeat={1}]
table {%
-30 0.0402930440035605
-20 0.358696256213143
-10 1.92375981728877
0.1 4.86252975188989
10 7.9329521051747
20 10.2067133811644
30 10.9408889011399
};

\addplot [semithick, myorange, mark=o, dashed, mark size=1.5, mark options={solid}, mark repeat={1}]
table {%
-30 0.0289641813086065
-20 0.253381975757357
-10 1.18368283225135
0.1 1.9893439740065
10 2.14471870414669
20 2.16236286588097
30 2.16414695636492
};

\addplot [semithick, myorange, mark=|,dashed, mark size=1.5, mark options={solid}, mark repeat={1}]
table {%
-30 0.0401261323170298
-20 0.353486173450936
-10 1.78674652552856
0.1 3.6819513284052
10 4.32672491421867
20 4.41681664457819
30 4.42619781822585
};

\addplot [semithick, myorange, mark=star, dashed,mark size=1.5, mark options={solid}, mark repeat={1}]
table {%
-30 0.0438754514922307
-20 0.386280960460996
-10 1.98906154807104
0.1 4.65572668798625
10 6.37596401890534
20 6.7937638373904
30 6.84407010105007
};

\addplot [semithick, myorange, mark=*,dashed, mark size=1.5, mark options={solid}, mark repeat={1}]
table {%
-30 0.0450003036807936
-20 0.395789827575198
-10 2.04141700609245
0.1 4.96892964658791
10 7.70476114540064
20 9.08685615319437
30 9.35412023464798
};


\addplot [semithick, myorange, mark=square , dashed, mark size=1.5, mark options={solid}, mark repeat={1}]
table{
-30 0.0454302275963866
-20 0.399309562501295
-10 2.057755121186
0.1 5.0558715624466
10 8.30449502590517
20 11.6222450436024
30 14.9437713040167
};

\addplot [thick, dashdotted, mypink,]
table {%
-30 0.00711705985720429
-20 0.0696097592777322
-10 0.577788597838845
0.1 2.56378403605454
10 5.33502159152586
20 7.20477875885344
30 7.68933680187468
};

\end{axis}

\end{tikzpicture}}
    \hspace{0.15\linewidth}
    \subfloat[$M=32$ , $b=1$, $K\in\{2, 4, 6\}$]{\label{fig:rsum_diff_k}
\begin{tikzpicture}[scale=0.8]

\definecolor{crimson2143940}{RGB}{214,39,40}
\definecolor{darkgray176}{RGB}{176,176,176}
\definecolor{darkorange25512714}{RGB}{255,127,14}
\definecolor{forestgreen4416044}{RGB}{44,160,44}
\definecolor{lightgray204}{RGB}{204,204,204}
\definecolor{steelblue31119180}{RGB}{31,119,180}
\definecolor{green01270}{RGB}{0,127,0}

\definecolor{myred}{HTML}{d62728}
\definecolor{myblue}{HTML}{1f77b4}
\definecolor{myolive}{HTML}{bcbd22}
\definecolor{mypurple}{HTML}{9467bd}
\definecolor{myorange}{HTML}{ff7f0e}
\definecolor{mypink}{HTML}{e377c2}
\definecolor{mygreen}{HTML}{2ca02c}
\definecolor{mybrown}{HTML}{8c564b}
\definecolor{mygray}{HTML}{7f7f7f}
\definecolor{mycyan}{HTML}{17becf}

\begin{axis}[
axis lines* = {left},
legend cell align={left},
legend style={nodes={scale=0.8, transform shape},
  fill opacity=0.8,
  draw opacity=1,
  text opacity=1,
  at={(0.03,0.97)},
  anchor=north west,
  draw=lightgray204
},
tick align=outside,
tick pos=left,
title={},
   x grid style={white!69.01960784313725!black},
        grid,
        grid style={help lines,color=gray!50, densely dotted},
        xlabel={$P_T/\sigma_v^2$ [dB]},
xmin=-33.25, xmax=38.25,
xtick style={color=black},
y grid style={darkgray176},
ylabel={$R_{\mathrm{sum}}$ [bits/channel use]},
ymin=-0.869273127070603, ymax=27.4946536076367,
ytick style={color=black}
]

\addlegendimage{semithick, myblue}
\addlegendentry{GNN}

\addlegendimage{semithick, dashed, myorange}
\addlegendentry{ZF}

\addlegendimage{only marks, black, mark=o, mark size=1.5, mark options={solid}, }
\addlegendentry{$K=2$}

\addlegendimage{only marks, black, mark=|, mark size=1.5, mark options={solid}, }
\addlegendentry{$K=4$}

\addlegendimage{only marks, black, mark=star, mark size=1.5, mark options={solid}, }
\addlegendentry{$K=6$}

\addplot [semithick, myblue, mark=o, mark size=1.5, mark options={solid}, mark repeat={1}]
table {%
-30 0.0127550383683584
-20 0.124924777138876
-10 1.04697746534634
0.1 4.69920450021582
10 9.13330682214005
20 10.9267071622
30 11.1979468887279
};

\addplot [semithick, myblue, mark=|, mark size=1.5, mark options={solid}, mark repeat={1}]
table {%
-30 0.0197747807674495
-20 0.194320839236532
-10 1.66827396526652
0.1 7.84286212007847
10 14.6166673978638
20 16.6365577305231
30 16.8926023598659
};

\addplot [semithick, myblue, mark=star, mark size=1.5, mark options={solid}, mark repeat={1}]
table {%
-30 0.0221647152262452
-20 0.21859209088121
-10 1.92987962224271
0.1 9.8177763590589
10 18.9287164418168
20 21.5675728726847
30 21.895749076106
};



\addplot [semithick, myorange, dashed, mark=o, mark size=1.5, mark options={solid}, mark repeat={1}]
table {%
-30 0.0279041269065782
-20 0.263568542718239
-10 1.743589616401
0.1 4.44788937368717
10 5.4175397855973
20 5.54981010074504
30 5.5635120796582
};

\addplot [semithick, myorange, mark=|,dashed, mark size=1.5, mark options={solid}, mark repeat={1}]
table {%
-30 0.0259337968372971
-20 0.252660215464046
-10 2.0289150193201
0.1 7.5917659625246
10 11.1328524281081
20 11.7676981095072
30 11.8366615930983
};

\addplot [semithick, myorange, mark=star,dashed, mark size=1.5, mark options={solid}, mark repeat={1}]
table {%
-30 0.0240654137806266
-20 0.236620267573259
-10 2.03552858482882
0.1 9.19867437010059
10 15.1538765100591
20 16.3895365830748
30 16.5279267915633
};





\end{axis}

\end{tikzpicture}}
    \caption{Achievable rates averaged over the channel realizations taken from the test set. Comparing the \gls{gnn} non-linear precoder against MRT or ZF, for the single-user case and a varying number of bits (a) and the multi-user case with one bit and a varying number of users (b).}
    \label{fig:singleuserplots}
\end{figure*}

\review{
For the following simulations, the total transmit power is normalized to $P_T=M$. During training, the \gls{snr} is fixed at $P_T/\sigma^2_v = 20$ dB. All models are trained for 20 epochs unless specified otherwise. The batch size is set to 128, the Adam optimizer~\cite{adam} is used with a learning rate of $5\times 10^{-3}$. During training, the loss function in~(\ref{eq:loss}) is computed numerically. Hence, per channel, $N_s = 125$ complex Gaussian symbols are generated per user according to $s_k \sim \mathcal{CN}(0,1)$. These symbols are precoded and quantized for the current channel using the \gls{gnn}. For this, non-uniform quantization levels are used, which are obtained using the Max-Lloyd algorithm, as described in~\cref{sec:non-uniform-quant}. Next, the average power constraint is enforced by performing a scalar normalization over these $N_s=125$ symbols according to \eqref{eq:norm}. Finally, these precoded and quantized symbols ($\vect{y}_{\mathrm{NL}}$), are used together with the intended transmit symbols ($\vect{s}$), to evaluate the expectation in the loss function according to~\eqref{eq:loss}. The training set consists of 200 000 generated channels, each of which is accompanied by $N_s=125$ symbol vectors. The channels are sampled from a complex normal distribution $[\mat{H}]_{i,j} \sim \mathcal{CN}(0,1)$ if a Rayleigh fading channel is assumed. However, when constructing radiation patterns, a \gls{los} channel is considered according to (\ref{eq:lineofsight}), where the user angle (in degrees) is randomly sampled from a discrete uniform distribution $\phi_k \sim \mathcal{U}\{0, 180\}$. All \gls{nn} hyperparameters are selected on a validation set of 1000 channel realizations, while the simulation results are obtained on an independent test set of 10 000 channel realizations.} The \gls{gnn} used in this section contains one input layer, four hidden layers ($N_h=4$), one output layer and $d_h=128$ features are learned in the hidden layers unless specified otherwise. The non-linear activation function in all hidden layers is the \gls{lrelu} which is defined as $\sigma(a) = \mathrm{max}(0, a)  - 0.01 \mathrm{min}(0,a)$. Finally, the temperature of the Gumbel-softmax estimator is set to $\tau=1$.

\subsection{GNN-Based Non-Linear Precoding Single-User Case}

In this section, the performance of the proposed GNN-based non-linear precoder is evaluated for the single-user case. In~\cref{fig:rsum_k1_b1} the rate is depicted for a varying \gls{snr} ($=P_T/\sigma_v^2$) for $M=32$ transmit antennas, $K=1$ user and a varying number of bits $b\in\{1, 2, 3, 4\}$. From this figure, it is clear that the \gls{gnn} outperforms \gls{mrt} precoding in the distortion-limited regime i.e., at high SNR. At high SNR, the GNN with one bit \glspl{dac} delivers slightly better performance than \gls{mrt} with three bit \glspl{dac}. This highlights the ability of the GNN-based precoder to leverage over-the-air combining to provide a higher-quality signal. Next to this, the GNN outperforms classical linear precoding for one to four bit \glspl{dac}. However, the performance gap decreases as the number of bits goes up. This is to be expected for two reasons: \textit{i)} as the number of bits rises, the distortion goes down, hence there is less to gain by using quantization-aware precoding  and \textit{ii)} the problem complexity increases, as the number of possible output levels rises exponentially with the number of bits, this increased search space results in a harder problem to solve. In conclusion, in the single-user case, the proposed method outperforms classical \gls{mrt} precoding for one to four bits. However, the gains are most pronounced in the one-bit case. 

\begin{figure*}[t!]
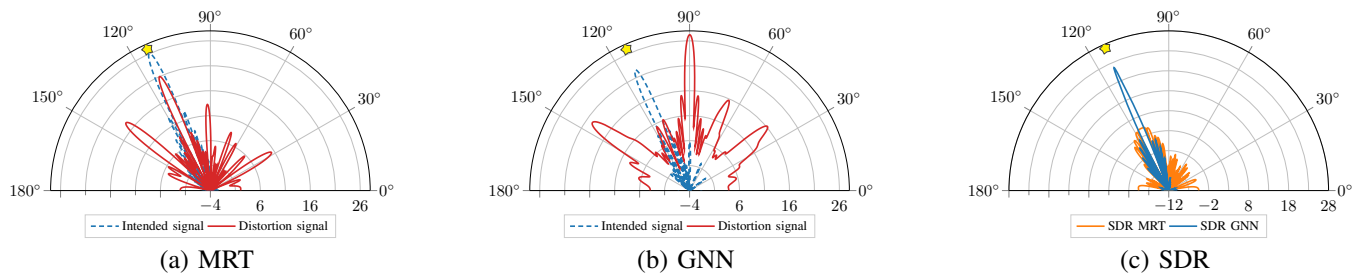

    \centering
    \subfloat[MRT]{\label{fig:rad_k1_mrt}\input{Figs/rad_MRT_k1_b1}}\hfill
    \subfloat[GNN]{\label{fig:rad_k1_gnn}\input{Figs/rad_GNN_k1_b1}}\hfill
     \subfloat[SDR]{\label{fig:rad_k1_sdr}\input{Figs/rad_SDR_k1_b1}}
	\caption{Radiation pattern of the intended signal $P_{\mathrm{lin}}(\Tilde{\phi})$ and distortion signal $
P_{\mathrm{dist}}(\Tilde{\phi})$ [dB] for a 1-bit DAC, $M=32$ and $K=1$. The \gls{gnn} precoder (a) and \gls{mrt} precoding (b) are compared for a pure \gls{los} channel and a half-wavelength \gls{ula}. The user is indicated by a star. In (c) the \acrfull{sdr} radiation pattern $P_{\mathrm{SDR}}(\Tilde{\phi})$ is depicted [dB]. }
	\label{fig:rad_k1_fixed_m} 
\end{figure*}

\begin{figure*}[ht!]
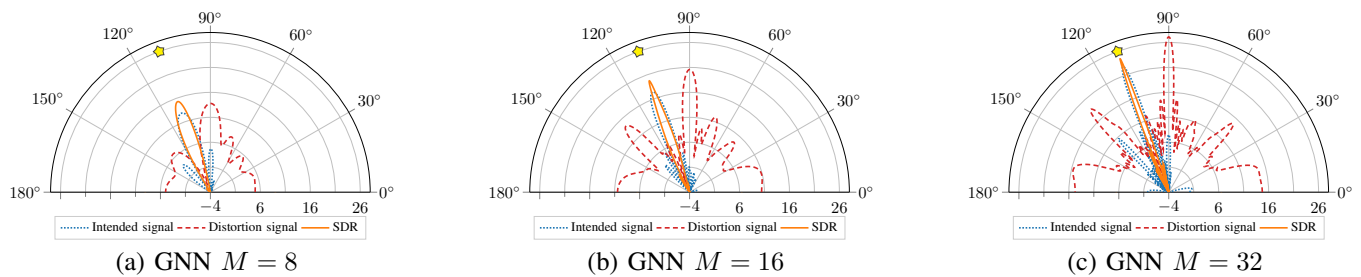

    \centering
    \subfloat[GNN $M=8$]{\label{fig:rad_M8}\input{Figs/rad_gnn_M8}}\hfill
    \subfloat[GNN $M=16$]{\label{fig:rad_M16}\input{Figs/rad_gnn_M16}}\hfill
     \subfloat[GNN  $M=32$]{\label{fig:rad_M32}\input{Figs/rad_gnn_M32}}
	\caption{Radiation pattern of the GNN for the intended signal $P_{\mathrm{lin}}(\Tilde{\phi})$ and distortion signal $
P_{\mathrm{dist}}(\Tilde{\phi})$ [dB] for a 1-bit DAC, varying number of transmit antennas $M \in \{8, 16, 32\}$ and $K=1$ user. The user angle is $\phi=110^{\circ}$. }
	\label{fig:rad_k1_gnn_varying_M} 
\end{figure*}

In~\cref{fig:rad_k1_fixed_m}, the radiation pattern of \gls{mrt} is compared to the \gls{gnn} for the one-bit single-user case $b=1$, $K=1$. This figure highlights that the \gls{gnn} achieves its gain by sending the distortion in non-user directions, while the distortion in the user direction is kept to a minimum. This results in a higher \gls{sdr} in the user direction as compared to \gls{mrt}, as can be seen in~\cref{fig:rad_k1_sdr}. In~\cref{fig:rad_k1_gnn_varying_M} the radiation pattern is depicted for the one-bit single-user case $b=1$, $K=1$ where the number of transmit antennas is varied $M \in \{8, 16, 32\}$. Given the increase in degrees of freedom when the number of transmit antennas grows, the GNN can control both the intended signal and distortion more precisely. This leads to an increase in \gls{sdr} in the user direction as the number of transmit antennas increases. 

\begin{figure*}[ht!]
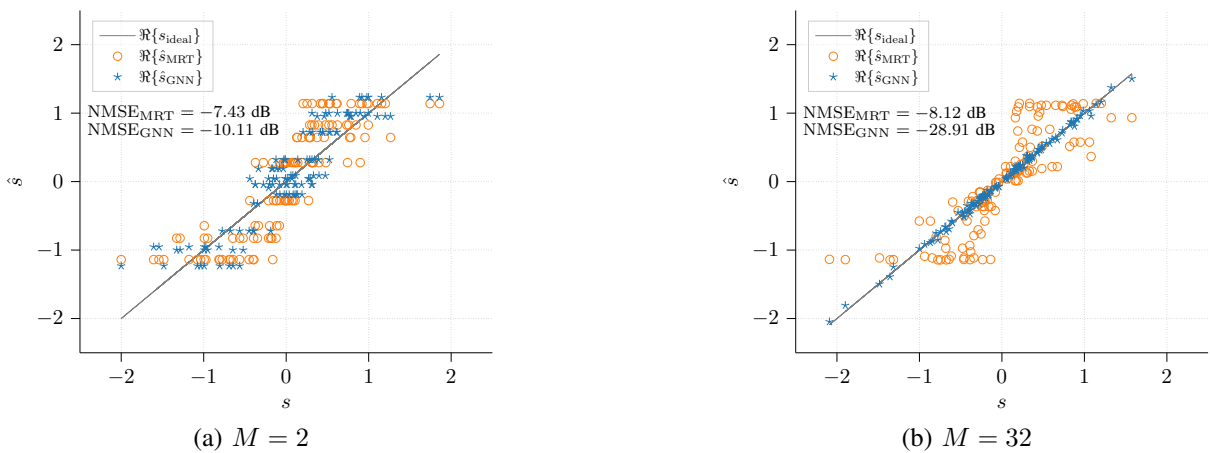

    \centering
    \subfloat[$M=2$]{\label{fig:shat_M2}\input{Figs/s_vs_shat/s_vs_shat_mrt_M2}}
    \hspace{0.15\linewidth}
    \subfloat[$M=32$]{\label{fig:shat_M32}\input{Figs/s_vs_shat/s_vs_shat_M32}}
    \caption{Real part of the estimated symbol $\hat{s}$ in function of the transited symbol $s$, over a single realization of a noiseless Rayleigh fading channel. Comparing MRT with GNN precoding, for $M=2$ (a) and $M=32$ (b), $K=1$ and $b=1$.}
    \label{fig:shat_global_M2}
\end{figure*}

In~\cref{fig:shat_M2} the transmitted symbols $s$ are plotted in function of the estimated received symbols $\hat{s}$ for $M=2$ transmit antennas, $K=1$ user and $b=1$ bit. In general, for complex channels and transmit symbols, after over-the-air combining, $2^{2bM}$ possible symbols can be received. These possible received signals are not uniformly distributed but depend on the channel realizations and the transmit signal at each antenna. Hence, selecting the right transmit signal at each antenna allows for control over the received signals. However, this is limited by the values of each channel realization. Nevertheless,~\cref{fig:shat_M2}, illustrates that the \gls{gnn} can more effectively utilize these $16$ possible symbols after over-the-air combining to obtain a good estimate close to the transmitted symbol. This is illustrated by the difference in \gls{nmse}, which is computed as $\mathrm{NMSE} = \frac{\expt{|s - \hat{s}|^2}}{\expt{|s|^2}}$. The \gls{gnn} achieves a \gls{nmse} of -10.11 dB, and MRT a \gls{nmse} of -7.43 dB. This increased performance scales with the number of transmit antennas as after over-the-air combining, $2^{2bM}$ possible symbols can be received. In~\cref{fig:shat_M32} the transmitted symbols $s$ are plotted in function of the estimated received symbols $\hat{s}$ for $M=32$ transmit antennas, $K=1$ user and $b=1$ bit \glspl{dac}. For \gls{mrt}, the estimated symbols differ largely from the transmitted symbols. When using \gls{gnn}-based non-linear precoding a remarkable improvement can be observed. This is expected given the increase in transmit antennas, which leads to $2^{2bM} = 2^{64}$ possible received symbols. This demonstrates the potential of using many transmit antennas in combination with one bit \glspl{dac}. However, the challenge lies in selecting the right combination of \gls{dac} output levels at each transmit antenna to obtain the desired received signal. \Cref{fig:shat_M32} demonstrates the ability of the \gls{gnn} to select the correct output levels at each \gls{dac} which results in estimated symbols that match very closely the transmit symbols. To quantify this, the \gls{nmse} is computed over all noiseless channel realizations and symbols from the test set. This is done for \gls{mrt} and \gls{gnn} precoding for $b\in\{1, 2, 3, 4\}$ bits in~\cref{tab:mse}. For one-bit \glspl{dac} the \gls{gnn} achieves a \gls{nmse} of -22.99 dB, while MRT leads to a \gls{nmse} of -5.38 dB, highlighting the performance gap between \gls{mrt} and the \gls{gnn}. As the number of bits increases, this performance gap decreases. However, even for $b=4$ bits, the GNN achieves a \gls{nmse} of -32.52 dB, while \gls{mrt} precoding achieves a \gls{nmse} of -27.70 dB. 

	\label{fig:rad_k1_mrt_varying_M} 

\begin{table}[t]
\centering
    \caption[Caption for LOF]{\gls{nmse} for MRT and GNN on all (noiseless) channels and symbols from the test set, for $M=32$, $K=1$.}
    \centering
 \begin{tabular}{@{}ccccc@{}} 
 \toprule
      Precoder & $b=1$  & $b=2$  & $b=3$ & $b=4$ \\ [0.5ex] 
\midrule
 \arrayrulecolor{black!30}
   NMSE MRT [dB] & -5.38 & -13.02 & -20.31 & -27.70  \\
\hline   

      NMSE GNN [dB] & -22.99 & -29.43 & -29.87 &  -32.52  \\

    \arrayrulecolor{black}
\bottomrule
\end{tabular}\label{tab:mse}
\end{table} 

\subsection{GNN-Based Non-Linear Precoding Multi-User Case}
In this section, the proposed precoder is evaluated in the multi-user scenario. In~\cref{fig:rsum_diff_k} the sum rate in function of the SNR is depicted for $M=32$ transmit antennas, $b=1$ bit and a varying number of users $K\in \{2, 4, 6\}$. This figure shows that the GNN outperforms \gls{zf} in the distortion-limited regime, i.e., at high SNR. \review{As the number of users increases, the distortion receives less of a beamforming gain, and as such becomes more spatially spread out. Consequently, there is less to gain in terms of system performance.} 
Additionally, this reduced gain is explained by the fact that the more users are present, the harder it is to train the \gls{gnn}. 

\begin{figure}
    \centering
\begin{tikzpicture}[scale=0.8]

\definecolor{crimson2143940}{RGB}{214,39,40}
\definecolor{darkgray176}{RGB}{176,176,176}
\definecolor{darkorange25512714}{RGB}{255,127,14}
\definecolor{forestgreen4416044}{RGB}{44,160,44}
\definecolor{lightgray204}{RGB}{204,204,204}
\definecolor{steelblue31119180}{RGB}{31,119,180}
\definecolor{green01270}{RGB}{0,127,0}

\definecolor{myred}{HTML}{d62728}
\definecolor{myblue}{HTML}{1f77b4}
\definecolor{myolive}{HTML}{bcbd22}
\definecolor{mypurple}{HTML}{9467bd}
\definecolor{myorange}{HTML}{ff7f0e}
\definecolor{mypink}{HTML}{e377c2}
\definecolor{mygreen}{HTML}{2ca02c}
\definecolor{mybrown}{HTML}{8c564b}
\definecolor{mygray}{HTML}{7f7f7f}
\definecolor{mycyan}{HTML}{17becf}

\begin{axis}[
axis lines* = {left},
legend cell align={left},
legend style={nodes={scale=0.8, transform shape},
  fill opacity=0.8,
  draw opacity=1,
  text opacity=1,
  at={(0.03,0.97)},
  anchor=north west,
  draw=lightgray204
},
tick align=outside,
tick pos=left,
title={},
   x grid style={white!69.01960784313725!black},
        grid,
        grid style={help lines,color=gray!50, densely dotted},
        xlabel={bits},
xmin=0.9, xmax=5.1,
xtick = {1, 2, 3, 4, 5},
xticklabels = {1, 2, 3, 4, $\infty$},
xtick style={color=black},
y grid style={darkgray176},
ylabel={$R_{\mathrm{sum}}$ [bits/channel use]},
ymin=-0.869273127070603, ymax=54.4946536076367,
ytick style={color=black}
]

\addlegendimage{semithick, myblue}
\addlegendentry{GNN $d_h=128$}

\addlegendimage{semithick, mypink}
\addlegendentry{GNN $d_h=256$}

\addlegendimage{semithick, dashed, myorange}
\addlegendentry{ZF}

\addlegendimage{semithick, dotted, myorange}
\addlegendentry{MRT}

\addplot [semithick, myblue, mark=o, mark size=1.5, mark options={solid}, mark repeat={1}]
table {%
1 7.25474323296877
2 9.3257330108142
3 9.58462843858504
4 10.2067133811644
};

\addplot [semithick, mypink, mark=o, mark size=1.5, mark options={solid}, mark repeat={1}]
table {%
3 10.4293838011449
4 10.6844925380207
};

\addplot [semithick, myblue, mark=|, mark size=1.5, mark options={solid}, mark repeat={1}]
table {%
1 10.9267071622
2 14.6379823363889
3 16.4800176849877
4 17.8058126563545
};

\addplot [semithick, mypink, mark=|, mark size=1.5, mark options={solid}, mark repeat={1}]
table {%
3 19.0089313750149
4  19.3363329338575
};

\addplot [semithick, myblue, mark=*, mark size=1.5, mark options={solid}, mark repeat={1}]
table {%
1 16.6365577305231
2 23.5618068318447
3 28.8608853899012
4 30.48700163135
};

\addplot [semithick, mypink, mark=*, mark size=1.5, mark options={solid}, mark repeat={1}]
table {%
3 30.7197304521327
4  33.7473268133004
};

\addplot [semithick, myblue, mark=star, mark size=1.5, mark options={solid}, mark repeat={1}]
table {%
1 21.5675728726847
2 31.4083230858039
3 36.9800369809415
4 40.3592326016706
};

\addplot [semithick, mypink, mark=star, mark size=1.5, mark options={solid}, mark repeat={1}]
table {%
3 39.8257366565965
4  46.1137636252063 
};

\addplot [semithick, myorange, dotted, mark=o, mark size=1.5, mark options={solid}, mark repeat={1}]
table {%
1 2.16236286588097
2 4.41681664457819
3  6.7937638373904
4 9.08685615319437
5 11.622 
};

\addplot [semithick, myorange, dashed, mark=|, mark size=1.5, mark options={solid}, mark repeat={1}]
table {%
1 5.54981010074504
2 10.3001043909686
3 14.719169324844
4 18.2258782180223
5 21.124 
};

\addplot [semithick, myorange, dashed, mark=*, mark size=1.5, mark options={solid}, mark repeat={1}]
table {%
1  11.7676981095072
2 20.155227226278
3 27.5980830070742
4 33.2552398613074
5 37.825 
};

\addplot [semithick, myorange, dashed, mark=star, mark size=1.5, mark options={solid}, mark repeat={1}]
table {%
1  16.3895365830748
2 28.0723757263054
3  38.4210935782809
4 46.2599569135317
5 52.584
};

\draw (3.5, 44.5) arc
[
    start angle=50, 
    end angle=310, 
    x radius=0.1cm, 
    y radius =0.5cm 
] ;
\draw[] (3.41, 42.5) -- (3.22, 44.5);
\node[] (k4) at (3.1, 45.5) {\footnotesize$K=6$};

\draw (3.5, 33.5) arc
[
    start angle=50,
    end angle=310,
    x radius=0.1cm,
    y radius =0.4cm
] ;
\draw[] (3.42, 33) -- (3.22, 34);
\node[] (k4) at (3.1, 35) {\footnotesize$K=4$};

\draw (3.5, 20.5) arc
[
    start angle=50,
    end angle=310,
    x radius=0.1cm,
    y radius =0.4cm
] ;
\draw[] (3.42, 20) -- (3.22, 22);
\node[] (k4) at (3.1, 23) {\footnotesize$K=2$};

\draw (3.5, 11.5) arc
[
    start angle=50,
    end angle=310,
    x radius=0.1cm,
    y radius =0.4cm
] ;
\draw[] (3.42, 6.2) -- (3.22, 4.2);
\node[] (k4) at (3.1, 3.2) {\footnotesize$K=1$};

\end{axis}

\end{tikzpicture}
    \caption{Achievable rates on test set, $P_T/\sigma_v^2 = 20$ dB. Comparing \gls{gnn} and MRT or \gls{zf} for $M=32$, $K\in\{1, 2, 4, 6\}$. A bigger GNN ($d_h=256$) is added for $b\in\{3, 4\}$.} 
    \label{fig:Rvsbits}
\end{figure}
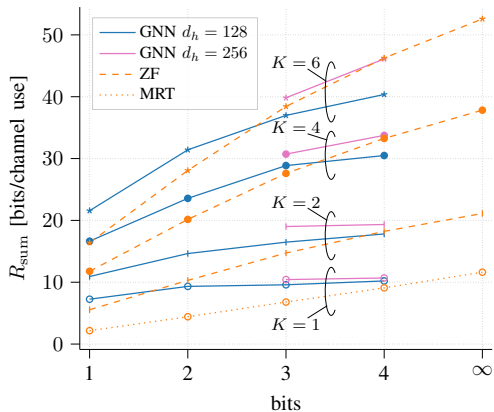

In~\cref{fig:Rvsbits}, the sum rate is illustrated in function of the number of bits for a fixed SNR$=P_T/\sigma_v^2=20$ dB, for $M=32$ transmit antennas and $K\in\{1, 2, 4, 6\}$ users. When comparing the GNN precoder with the baseline (\gls{mrt} or \gls{zf}) it is clear that the proposed method is most performant when few users are present and few bits are considered. When more users are present and/or more bits are considered, the proposed method shows similar or lower performance than the baseline. This is expected as the potential gains diminish as the number of users increases, and as the complexity of the problem exponentially increases with the number of bits. Note that in~\cref{fig:Rvsbits} two \glspl{gnn} are considered, the standard model with $d_h=128$ features and a bigger model with $d_h=256$ features. When the problem complexity is higher, i.e., more bits and users are considered, it is shown that the larger \gls{gnn} ($d_h=256$) achieves better performance as compared to the smaller model ($d_h=128$) and \gls{zf} precoding. However, this comes at the cost of an increased model size and complexity. Nevertheless, previously the most critical case was identified as the one with few users and few bits, in this regime, the proposed smaller model ($d_h=128$) outperforms the baseline.


\section{Power Consumption} \label{sec:power_cons}
To quantify the power consumption of the proposed precoder, the consumed power of the \glspl{dac} and the processing due to the \gls{gnn} is taken into account.

\subsection{DACs Power Consumption}
We consider the widely used current steering \gls{dac} architecture, with the following power consumption model~\cite{pwr_dac}
\begin{align}
    P_{\mathrm{DAC}} \approx \frac{1}{2} V_{dd} I_0 (2^b - 1) + b C_p \frac{f_s}{2} V_{dd}^2 \label{eq:daccons}
\end{align}
where $V_{dd}$ is the supply voltage, $I_0$ the current of the current source of the least significant bit and $C_p$ the parasitic capacitance of each transistor. Based on~\cite{pwr_dac} we use $V_{dd} =$ \SI{3}{\volt}, $I_0=$ \SI{10}{\micro\ampere} and $C_p=$ \SI{1}{\pico \farad}.

For a fair comparison of the consumed power, a fixed quality of service is selected at which the proposed precoder is compared to the conventional precoder. As a concrete example, a rate constraint of 7 bits/channel use is considered for the scenario in~\cref{fig:rsum_k1_b1} at an \gls{snr} = \SI{20}{\decibel}. From this figure, it is clear that the \gls{gnn} can meet this constraint by using \glspl{dac} with $b=1$ bits while \gls{mrt} requires $b=3$ bits. Given the required bit resolution and sampling frequency, the consumed power of all \glspl{dac} is $P_{\mathrm{DACs, tot}} = 2 M P_{\mathrm{DAC}}$. \review{Note that we here consider two types of \glspl{dac}, namely the classical baseband \gls{dac}, whereby the system relies on analog upconversion to place the signal at the desired carrier frequency and the \gls{rfdac} which directly synthesizes the signal at the desired carrier frequency. For the classical baseband \gls{dac} we consider a sampling frequency of $f_s =  4 B$, where $B$ is the bandwidth and an oversampling factor of four is applied. For the \gls{rfdac} we select $f_s = 4f_c / (2n-1)$, which places the baseband signal at the center of the $n^{\mathrm{th}}$ Nyquist zone. In practice, the second Nyquist zone ($n=2$) is most commonly utilized~\cite {rfsoc_book}. Note that for the \gls{rfdac} the sampling frequency is independent of the bandwidth, however, the condition $f_s \geq B$ must be satisfied, which is generally the case, hence $f_s = \frac{4}{3}f_c$. The consumption of the baseband \glspl{dac} is illustrated in~\cref{fig:dac_cons} in function of the bandwidth $B$. The proposed precoder allows the \glspl{dac} to consume a factor of 4-7 less power. In~\cref{fig:rfdac_cons} the power consumption of the \glspl{rfdac} is given as a function of the bandwidth. Note that this consumption is constant over the bandwidth as the sampling frequency for the \gls{rfdac} is linked to the carrier frequency $f_c=3.5$ GHz, rather than the bandwidth. The proposed precoder allows the \glspl{rfdac} to consume a factor of 3 less power as compared to \gls{mrt}. This is to be expected as the power consumption of the \glspl{rfdac} is dominated by the dynamic consumption (term 2 in~\eqref{eq:daccons}), which scales linearly with the number of bits, while the sampling frequency is fixed for both precoders.}

\subsection{GNN Processing Power Consumption}
As computed in~\cref{sec:complexity}, the total number of \gls{flops} required to run the \gls{gnn} one time is $\mathcal{O}_{\mathrm{GNN}}^{(\mathrm{add})} + \mathcal{O}_{\mathrm{GNN}}^{(\mathrm{mul})}$. However, given that non-linear precoding is considered, the \gls{gnn} needs to be executed for each transmit symbol. Hence, to compute how many forward passes are required per second, the symbol rate is computed. Consider a transmission bandwidth $B=(1+\alpha_{\mathrm{rol}})/T$, where $T$ is the symbol period and $\alpha_{\mathrm{rol}}=0.1$ a roll-off factor. This gives a symbol rate of $R_s =  \frac{B}{1 + \alpha_{\mathrm{rol}}}$ symbols per second. Given this, the GNN needs to precode $R_s$ symbols per second. Hence the total time complexity of the GNN is $R_s (\mathcal{O}_{\mathrm{GNN}}^{(\mathrm{add})} +  \mathcal{O}_{\mathrm{GNN}}^{(\mathrm{mul})})$ \gls{flops} per second. Given this, we can compute the required speed of the accelerator given a certain bandwidth. For instance, when considering $B=$ \SI{1}{\mega\hertz} the considered accelerator requires a speed of 12 T\gls{flops}/second. 
Based on the \gls{nn} accelerator comparison from~\cite{NN_HW_comparison}, we select the accelerator from~\cite{HW_accelerator} that achieves the speed requirement of 12 T\gls{flops}/s and has a high efficiency of $\eta =646.6$ TFLOPs/s/W. This accelerator operates on 8-bit floating point numbers while the current \gls{gnn} is trained on 32-bit floating point numbers. The influence of this reduced precision on the performance of the \gls{gnn} should be further investigated. Nevertheless, to get an estimate of the achievable power consumption of the \gls{gnn} processing, it is expressed in function of the bandwidth as $P_{\mathrm{GNN}} = \frac{1}{\eta} \frac{B}{1 + \alpha_{\mathrm{rol}}} \left(\mathcal{O}_{\mathrm{GNN}}^{(\mathrm{add})} + \mathcal{O}_{\mathrm{GNN}}^{(\mathrm{mul})}\right)$. First we consider the baseband \glspl{dac}, from~\cref{fig:dac_cons} it can be noted that the proposed solution provides a power reduction when the system bandwidth is below \SI{200}{\kilo\hertz}. When the bandwidth is further increased, the gain of using fewer bits in the \glspl{dac} is overshadowed by the increased \gls{gnn} processing power consumption. However, by noting that the computational complexity of the \gls{gnn} in~\eqref{eq:compl_mul} and~~\eqref{eq:compl_add} scales linearly with the number of layers $N_h$ but quadratically with the number of features $d_h$, the processing power consumption can be reduced. By, changing the depth and width of the \gls{gnn} to $N_h=8$, $d_h=32$ rather than $N_h=4$, $d_h=128$, the performance (for $K=1$, $b=1$) is maintained while the power consumption is drastically reduced, as can be seen in~\cref{fig:rsum_k1_b1} and~\cref{fig:dac_cons}. This results in a power reduction for a system bandwidth up to \SI{3.5}{\mega\hertz}. Note that the speed requirement for the \gls{gnn} with $N_h=8$, $d_h=32$ is still met, as the selected accelerator from~\cite{HW_accelerator} has a maximum speed of 39.8 T\gls{flops}/s, while at $B=$ \SI{4}{\mega\hertz} the \gls{gnn}, requires 8.82 T\gls{flops}/s. 
\review{Next to this, we study the power consumption when considering \glspl{rfdac} in~\cref{fig:rfdac_cons}. The consumption of \glspl{rfdac} is much larger as compared to baseband \glspl{dac}, hence the added consumption of the \gls{gnn} processing is minor as compared to the \gls{rfdac} consumption. This leads to power savings, which are maintained for much higher bandwidths. However, in this situation, the limiting factor is the speed of the \gls{nn} accelerator rather than its power consumption. The accelerator from~\cite{HW_accelerator} has a maximum speed of 39.8 T\gls{flops}/s. This limits the maximum bandwidth when using the \gls{gnn} with $N_h=4$, $d_h=128$ to 1.95 MHz\footnote{Note that for completeness the expected power consumption for higher bandwidths is also added to~\cref{fig:rfdac_cons}.} while the maximum achievable bandwidth for the \gls{gnn} with $N_h=8$, $d_h=32$ is 15.8 MHz. Nevertheless, over the 15.8 MHz bandwidth, a power reduction of 2.9 is maintained.} As a perspective, future work could investigate more in-depth the performance and power consumption trade-off for the optimal \gls{gnn} structure.


Additionally, it is worth noting that, the current analysis only considers the increase in \gls{gnn} power consumption and the decrease in \glspl{dac} power consumption, while other factors might be at play. For instance, the reduced resolution could have auxiliary effects as the power consumption in the digital chain can be reduced when this lower resolution is exploited, e.g., the power consumption of digital filters can be reduced as fewer bits need to be computed. Moreover, the power consumption of the digital fronthaul, which connects the baseband unit with the antenna array, can also be reduced when fewer bits are required. This could allow a large energy reduction, especially in a distributed/cell-free scenario where long fronthaul connections may connect central processing units to access points. These indirect effects should be considered in a further analysis.



\begin{figure}
    \centering
\begin{tikzpicture}[scale=0.8]

\definecolor{crimson2143940}{RGB}{214,39,40}
\definecolor{darkgray176}{RGB}{176,176,176}
\definecolor{darkorange25512714}{RGB}{255,127,14}
\definecolor{forestgreen4416044}{RGB}{44,160,44}
\definecolor{lightgray204}{RGB}{204,204,204}
\definecolor{steelblue31119180}{RGB}{31,119,180}
\definecolor{green01270}{RGB}{0,127,0}

\definecolor{myred}{HTML}{d62728}
\definecolor{myblue}{HTML}{1f77b4}
\definecolor{myolive}{HTML}{bcbd22}
\definecolor{mypurple}{HTML}{9467bd}
\definecolor{myorange}{HTML}{ff7f0e}
\definecolor{mypink}{HTML}{e377c2}
\definecolor{mygreen}{HTML}{2ca02c}
\definecolor{mybrown}{HTML}{8c564b}
\definecolor{mygray}{HTML}{7f7f7f}
\definecolor{mycyan}{HTML}{17becf}

\begin{axis}[
axis lines* = {left},
legend cell align={left},
legend style={nodes={scale=0.8, transform shape},
  fill opacity=0.8,
  draw opacity=1,
  text opacity=1,
  at={(0.03,0.97)},
  anchor=north west,
  draw=lightgray204
},
tick align=outside,
tick pos=left,
title={},
   x grid style={white!69.01960784313725!black},
        grid,
        grid style={help lines,color=gray!50, densely dotted},
        xlabel={Bandwidth [MHz]},
xmin=-0.1, xmax=4.51,
xtick style={color=black},
y grid style={darkgray176},
ylabel={Power [mW]},
ymin=-0.07, ymax=37.64,
ytick style={color=black}
]
\addplot [semithick, myblue, mark=o, mark size=2, mark options={solid}, mark repeat={1}]
table {%
0.001 0.961152
0.501 1.537152
1.001 2.113152
1.501 2.689152
2.001 3.265152
2.501 3.841152
3.001 4.417152
3.501 4.993152
4.001 5.569152
};
\addlegendentry{$P_{\mathrm{DACs, tot, GNN}}$ ($b=1$)}
\addplot [semithick, myorange, mark=*, mark size=2, mark options={solid}, mark repeat={1}]
table {%
0.001 6.723456
0.501 8.451456
1.001 10.179456
1.501 11.907456
2.001 13.635456
2.501 15.363456
3.001 17.091456
3.501 18.819456
4.001 20.547456
};
\addlegendentry{$P_{\mathrm{DACs, tot, MRT}}$ ($b=3$)}


\addplot [semithick, myred, mark=star, mark size=2, mark options={solid}, mark repeat={1}]
table {%
0.001 0.992803412985406
0.501 17.3945099056885
1.001 33.7962163983916
1.501 50.1979228910947
2.001 66.5996293837977
2.501 83.0013358765008
3.001 99.4030423692039
3.501 115.804748861907
4.001 132.20645535461
};
\addlegendentry{$P_{\mathrm{GNN}} + P_{\mathrm{DACs, tot, GNN}}$ ($b=1$, $N_h=4$, $d_h=128$)}

\addplot [semithick, mygray, mark=|, mark size=2, mark options={solid}, mark repeat={1}]
table {%
0.001 0.965048859095127
0.501 3.48947840665861
1.001 6.01390795422208
1.501 8.53833750178556
2.001 11.062767049349
2.501 13.5871965969125
3.001 16.111626144476
3.501 18.6360556920395
4.001 21.160485239603
};
\addlegendentry{$P_{\mathrm{GNN}} + P_{\mathrm{DACs, tot, GNN}}$ ($b=1$, $N_h=8$, $d_h=32$)}

\end{axis}
\end{tikzpicture}
    \caption{Power consumption of the \glspl{dac} and \gls{gnn} processing in function of the bandwidth.}
    \label{fig:dac_cons}
\end{figure}

\begin{figure}
    \centering
\begin{tikzpicture}[scale=0.9]

\definecolor{crimson2143940}{RGB}{214,39,40}
\definecolor{darkgray176}{RGB}{176,176,176}
\definecolor{darkorange25512714}{RGB}{255,127,14}
\definecolor{forestgreen4416044}{RGB}{44,160,44}
\definecolor{lightgray204}{RGB}{204,204,204}
\definecolor{steelblue31119180}{RGB}{31,119,180}
\definecolor{green01270}{RGB}{0,127,0}

\definecolor{myred}{HTML}{d62728}
\definecolor{myblue}{HTML}{1f77b4}
\definecolor{myolive}{HTML}{bcbd22}
\definecolor{mypurple}{HTML}{9467bd}
\definecolor{myorange}{HTML}{ff7f0e}
\definecolor{mypink}{HTML}{e377c2}
\definecolor{mygreen}{HTML}{2ca02c}
\definecolor{mybrown}{HTML}{8c564b}
\definecolor{mygray}{HTML}{7f7f7f}
\definecolor{mycyan}{HTML}{17becf}

\begin{axis}[
axis lines* = {left},
legend cell align={left},
legend style={nodes={scale=0.8, transform shape},
  fill opacity=0.8,
  draw opacity=1,
  text opacity=1,
  at={(0.03,0.93)},
  anchor=north west,
  draw=lightgray204
},
tick align=outside,
tick pos=left,
title={},
   x grid style={white!69.01960784313725!black},
        grid,
        grid style={help lines,color=gray!50, densely dotted},
        xlabel={Bandwidth [MHz]},
xmin=-0.1, xmax=16.3,
xtick style={color=black},
y grid style={darkgray176},
ylabel={Power [mW]},
ymin=1210, ymax=4175.928,
ytick style={color=black}
]
\addplot [semithick, myblue, mark=o, mark size=2, mark options={solid}, mark repeat={1}]
table {%
0.001 1344.96
0.816736842105263 1344.96
1.63247368421053 1344.96
2.44821052631579 1344.96
3.26394736842105 1344.96
4.07968421052632 1344.96
4.89542105263158 1344.96
5.71115789473684 1344.96
6.52689473684211 1344.96
7.34263157894737 1344.96
8.15836842105263 1344.96
8.9741052631579 1344.96
9.78984210526316 1344.96
10.6055789473684 1344.96
11.4213157894737 1344.96
12.2370526315789 1344.96
13.0527894736842 1344.96
13.8685263157895 1344.96
14.6842631578947 1344.96
15.5 1344.96
};
\addlegendentry{$P_{\mathrm{RFDACs, tot, GNN}}$ ($b=1$)}
\addplot [semithick, myorange, mark=*, mark size=2, mark options={solid}, mark repeat={1}]
table {%
0.001 4038.72
0.816736842105263 4038.72
1.63247368421053 4038.72
2.44821052631579 4038.72
3.26394736842105 4038.72
4.07968421052632 4038.72
4.89542105263158 4038.72
5.71115789473684 4038.72
6.52689473684211 4038.72
7.34263157894737 4038.72
8.15836842105263 4038.72
8.9741052631579 4038.72
9.78984210526316 4038.72
10.6055789473684 4038.72
11.4213157894737 4038.72
12.2370526315789 4038.72
13.0527894736842 4038.72
13.8685263157895 4038.72
14.6842631578947 4038.72
15.5 4038.72
};
\addlegendentry{$P_{\mathrm{RFDACs, tot, MRT}}$ ($b=3$)}


\addplot [semithick, myred, mark=star, mark size=2, mark options={solid}, mark repeat={1}]
table {%
0.001 1344.99165141299
0.816736842105263 1370.81087508987
1.63247368421053 1396.63009876676
2.44821052631579 1422.44932244364
3.26394736842105 1448.26854612052
4.07968421052632 1474.08776979741
4.89542105263158 1499.90699347429
5.71115789473684 1525.72621715118
6.52689473684211 1551.54544082806
7.34263157894737 1577.36466450495
8.15836842105263 1603.18388818183
8.9741052631579 1629.00311185872
9.78984210526316 1654.8223355356
10.6055789473684 1680.64155921249
11.4213157894737 1706.46078288937
12.2370526315789 1732.28000656626
13.0527894736842 1758.09923024314
13.8685263157895 1783.91845392003
14.6842631578947 1809.73767759691
15.5 1835.5569012738
};
\addlegendentry{$P_{\mathrm{GNN}} + P_{\mathrm{RFDACs, tot, GNN}}$ ($b=1$, $N_h=4$, $d_h=128$)}

\addplot [semithick, mygray, mark=|, mark size=2, mark options={solid}, mark repeat={1}]
table {%
0.001 1344.9638968591
0.816736842105263 1348.14270839148
1.63247368421053 1351.32151992387
2.44821052631579 1354.50033145626
3.26394736842105 1357.67914298865
4.07968421052632 1360.85795452104
4.89542105263158 1364.03676605342
5.71115789473684 1367.21557758581
6.52689473684211 1370.3943891182
7.34263157894737 1373.57320065059
8.15836842105263 1376.75201218298
8.9741052631579 1379.93082371536
9.78984210526316 1383.10963524775
10.6055789473684 1386.28844678014
11.4213157894737 1389.46725831253
12.2370526315789 1392.64606984492
13.0527894736842 1395.8248813773
13.8685263157895 1399.00369290969
14.6842631578947 1402.18250444208
15.5 1405.36131597447
};
\addlegendentry{$P_{\mathrm{GNN}} + P_{\mathrm{RFDACs, tot, GNN}}$ ($b=1$, $N_h=8$, $d_h=32$)}

\end{axis}
\end{tikzpicture}
    \caption{Power consumption of the \glspl{rfdac} and \gls{gnn} processing in function of the bandwidth.}
    \label{fig:rfdac_cons}
\end{figure}

\review{
\section{Practical Implementation Challenges}
While the proposed methods are promising in terms of system-level performance, practical deployment raises several challenges. Given that the proposed method relies on non-linear precoding, the computational complexity is linked to the symbol rate and hence, the system bandwidth. Consequently, the current system is limited to bandwidths of 15.8 MHz due to the limited processing speed of the current state-of-the-art hardware accelerator. However, given the rapid advancements in hardware accelerators driven by growing demand for \gls{nn} applications, the speed and efficiency of these hardware accelerators are expected to improve further~\cite{NN_HW_comparison}. Furthermore, training and deployment raise several challenges. The base model could be trained on simulation data (generated channels in our case). However, the distribution of the real-life measured channels will be different from the simulated channels. In the literature, this phenomenon is known as covariate shift~\cite{tianzheng_finetuning}. Even so, by training in a simulated environment, a good initialization point of the model can be obtained, which can then be further fine-tuned using a limited set of measured real-life data~\cite{tianzheng_finetuning}. By pre-training the model on simulation data, the amount of real-life measurements, which are often expensive and time-consuming to collect, can be limited~\cite{tianzheng_finetuning}. Once deployed, another challenge arises, namely, maintaining the model when the data distribution varies over time. For instance, the properties of the wireless environment can change due to changes in the physical environment. Hence, the model needs to be continually updated. However, a trade-off arises here to avoid the problem known as catastrophic forgetting, where updating the model on new data usually results in a severe performance degradation on the old data~\cite{continual_learning}. To overcome these challenges, we refer to the rich literature on continual learning~\cite{continual_learning}. Finally, the proposed models should be compliant with current standards. As it stands, machine learning models are not standardized~\cite{ai_standardization}. However, testability and consistent system behavior are concerns. Hence, these models should be extensively tested to meet current standards. More specifically, related to this work practical constraints such as the \gls{aclr} might be problematic due to the non-linear behavior of the \glspl{dac}. Such constraints could be incorporated into the loss function, during training, to be compliant with the current standard. 
}
\FloatBarrier

\section{Conclusion}
\label{sec:conclusion}
In this work, a coarsely quantized downlink massive \gls{mimo} system is considered. A \gls{gnn} is proposed which leverages the many antennas at the \gls{bs} to transmit the non-linear quantization distortion in non-user directions. By doing so, a significant increase in achievable sum rate is obtained. For instance, when considering one-bit quantization in the single user case at high \gls{snr}, a three times increase in achievable rate as compared to classical precoding is obtained.  This allows for using \glspl{dac} with fewer bits to achieve a similar sum rate as traditional precoding techniques, which require more bits. \review{The power consumption of baseband \glspl{dac} is reduced by a factor of 4-7. However, when considering the additional processing power consumption, the total power reduction is only valid for bandwidths up to \SI{3.5}{\mega\hertz}. Additionally, the power consumption of \glspl{rfdac} is reduced by a factor of 3. However, due to their higher absolute power consumption, the impact of the additional processing power consumption is limited, as the total power consumption is still reduced by a factor of 2.9.} Next to this, a more holistic power consumption analysis, including the full digital chain and fronthaul consumption, can further improve these results. Moreover, future work should address the high processing power. For this two options remain open. \textit{i)} The proposed solution can be further optimized by studying smaller \gls{gnn} architectures, combined with model pruning and quantization of the model weights and activations. By doing so, the computational complexity of the current model can be significantly reduced. \textit{ii)} The proposed solution relies on non-linear precoding. As such, the computational complexity scales linearly with the system bandwidth, which is the critical issue concerning the high processing power. Although challenging, future work could investigate this issue by using linear precoding and/or reduced-complexity non-linear precoders, which significantly reduce the computational complexity. \review{Finally, the impact of channel estimation errors on the performance of the proposed method should be investigated.}


\linespread{0.9}
\bibliographystyle{IEEEtran}
\bibliography{References}

\end{document}